\documentclass[aps,pre,twocolumn,superscriptaddress,longbibliography]{revtex4-2}
\usepackage[T1]{fontenc}
\usepackage{graphicx}
\usepackage{amsmath, amssymb}
\usepackage{commath}
\usepackage{xcolor}
\usepackage{hyperref}
\begin{document}

\title{Design principles for energy dissipation in viscoelastic network metamaterials}

\author{Niranjan Sarpangala}
\affiliation{Department of Physics and Astronomy, University of Pennsylvania, Philadelphia, PA}

\author{Sean Fancher}
\affiliation{Department of Physics and Astronomy, University of Pennsylvania, Philadelphia, PA}
\affiliation{Department of Biophysics, University of Michigan, Ann Arbor, MI}
\author{Prashant K. Purohit}
\affiliation{Department of Mechanical Engineering and Applied Mechanics, University of Pennsylvania, Philadelphia, PA}

\author{Eleni Katifori}
\affiliation{Department of Physics and Astronomy, University of Pennsylvania, Philadelphia, PA}
\affiliation{Center for Computational Biology, Flatiron institute, New York, NY}
\begin{abstract}
Mechanical energy dissipation in networked materials is relevant for applications from vibration isolation to impact protection, yet identifying optimal dissipative architectures in large disordered truss networks is computationally prohibitive with conventional finite element methods. We develop an efficient graph Laplacian-based spectral framework for viscoelastic truss networks, in which the full continuum dynamics of each rod are retained exactly and the problem size scales with the number of joints rather than element-level discretization points. Using this framework, we investigate how redistributing cross-sectional areas within a network (without changing material composition) controls energy dissipation. We find that random redistribution typically reduces dissipation relative to a uniform baseline, while gradient-based optimization yields nontrivial architectures whose form is governed by the intrinsic attenuation length of the base material. Focusing on driving frequencies near a global resonant mode of the network, we show that the optimal mass distribution decays from the source (driven joint) with the attenuation length scale, and at small attenuation lengths the optimal architecture is independent of the boundary conditions. These results motivate future studies of dissipation length scale based design principles on more complex disordered architectures and provide an efficient computational framework for exploring such structures at scale.
\end{abstract}

\maketitle
\section{Introduction}
Metamaterials are broadly described as materials whose effective properties arise from the rational design of their underlying architecture rather than solely from their chemical composition \cite{wegener2013metamaterials,zheludev2010road,yu2018mechanical}. 
Extensive research has led to mechanical metamaterials for applications including seismic protection and vibration isolation \cite{wu2024wave}, cloaking \cite{wang2022mechanical}, auxetic response \cite{reid2018auxetic}, and energy focusing \cite{bordiga2024automated}. 
While some of these functionalities relate to the material’s ability to dissipate energy, it was also shown that dissipation by the base material could be harnessed to obtain new mechanical functionalities that are not easy to obtain in lossless systems \cite{dykstra2022extreme,jia2020engineering}.

The ability of materials to damp vibrations or dissipate mechanical energy is crucial to applications such as earthquake protection \cite{spencer2003state}, vibration isolation \cite{al2022advances}, and packaging\cite{zhang2023mechanical}. 
Traditionally, energy dissipation has been achieved using porous materials, foams \cite{sun2018dynamic, banhart1996damping} or composites \cite{strkek2021highly}; however, these materials have less tunability \cite{al2022advances}. 
In contrast, precise networked structures \cite{schaedler2016architected, jenett2020discretely, meza2015resilient, reid2018auxetic} where physical properties of edges, network geometry and topology could be independently altered, thanks to advances in additive manufacturing, open a large tunable design space for mechanical functionalities, which could also be applied to the mechanical function of energy dissipation \cite{al2022advances,fayyaz2025damping, li2025comprehensive}. Designs that use a single constituent material may also simplify recycling compared to certain composite systems.

Mechanical energy dissipation can arise from multiple mechanisms, including frictional damping \cite{jeong2025energy} and electromagnetic dissipation \cite{oz2025experimental}. 
In this work, we focus on linear viscoelasticity, which is particularly relevant for additively manufactured polymeric materials. This mechanism allows for repeated use over many loading cycles, and can thus offer longer operational lifetimes than friction-based mechanisms. Moreover, linear viscoelastic materials are well characterized \cite{lakes2009viscoelastic}, making them a suitable platform for developing network-level theoretical descriptions. However, general principles for designing spatial architectures of linear viscoelastic materials with desired properties, in particular related to energy dissipation are unclear.

The class of problems concerned with finding optimal material distributions under prescribed mechanical objectives is the subject of topology optimization, a well-established framework in solid mechanics~\cite{sigmund2013topology}. While substantial work in this 
direction has addressed dissipative systems~\cite{zhang2020topology,gupta2025additive}, these are typically based on finite-element discretization and 
focus on designs of continuum or periodic-lattice geometries. Their extension 
to large, disordered network architectures remains computationally 
demanding and hence underexplored.


Mechanical dissipation is also important in biological context, in particular for protective biological tissues, including  bone~\cite{rudenko2016skeletal,bailey2018mechanical} and in particular, woodpecker skulls~\cite{yoon2011mechanical}. Flow dissipation also plays a functional role in biological transport 
networks such as leaf venation~\cite{katifori2010damage}, and blood vasculature \cite{chatterjee2026hierarchical}. Hierarchical loopy architectures in leaf venation have been explained through dissipation-minimizing edge conductivity 
optimization under a material cost constraint that are robust under damage and fluctuations ~\cite{katifori2010damage,ronellenfitsch2016global}. A simpler optimization to minimize dissipation under a material cost constraint yields tree-like and gradient architectures in conductivities~\cite{durand2007structure}. Since flow and mechanical networks are formally analogous in the linear regime~\cite{firestone1933new, ortiz2025unified}, it is interesting to explore whether dissipative mechanical networks admit a comparable design principle.

We first address the question of exploring the large design space in viscoelastic truss networks efficiently in computation, by extending a graph Laplacian based approach we previously developed \cite{fancher2025analysis} to dissipative systems. In this approach, we treat each rod as a linear viscoelastic element and solve the constitutive equation together with the equation of motion for uniaxial stress along rods to get an analytical solution of internal stress in rods. These solutions are then used for force balance at network nodes and construct a linear relation between Fourier transformed forces and displacement of rods. This matrix that is of the size of number of nodes (much smaller than matrix sizes in FEM) is then used to perform gradient-descent based optimization of rod cross sectional areas to maximize total dissipated energy under a fixed material cost for triangular networks driven by harmonic excitation.

Prior work on linear viscoelastic materials has established that energy dissipation depends on geometry, but these studies have been largely confined to periodic lattice architectures \cite{welander2025tailored,fayyaz2025damping}. Here, we investigate how redistributing cross-sectional areas across network edges (without changing material composition) affects total dissipation in viscoelastic trusses under harmonic excitation. We find that the optimal architectures that maximize energy dissipation are governed by the intrinsic attenuation length of the base material. The characteristic decay length of the optimal mass distribution depends on the attenuation length of the base material. Deviations from this scaling arise from finite network size and discrete edge length effects. We also observe that at small attenuation lengths, the optimal architecture is independent of imposed boundary conditions at domain walls far from source. Together, these results suggest attenuation-length-governed design principles for linear dissipative metamaterials, and motivate their extension to more general disordered network architectures.

The remainder of the paper is organized as follows. Section~\ref{truss_formalism} presents the viscoelastic network Laplacian framework and analyzes dissipation length scales in bulk materials. Section~\ref{sec:optimization} presents benchmarking, random network results, and gradient-based optimization. We conclude with design principles and outlook in 
Section~\ref{sec:conclusion}.

\section{Viscoelastic truss network model}
\label{truss_formalism}
In our previous work on purely elastic truss networks, we developed a graph Laplacian-based spectral framework in temporal frequency domain for truss networks that captures the uniaxial rod dynamics through a network Laplacian, that relates joint forces to joint displacements in the frequency domain \cite{fancher2025analysis}. This approach retains the uniaxial continuum behavior of each rod, avoids element-level discretization, and provides an exact continuum limit of mass–stiffness finite element models for small amplitude dynamics.\\

\begin{figure}
    \centering
    \includegraphics[width=\linewidth]{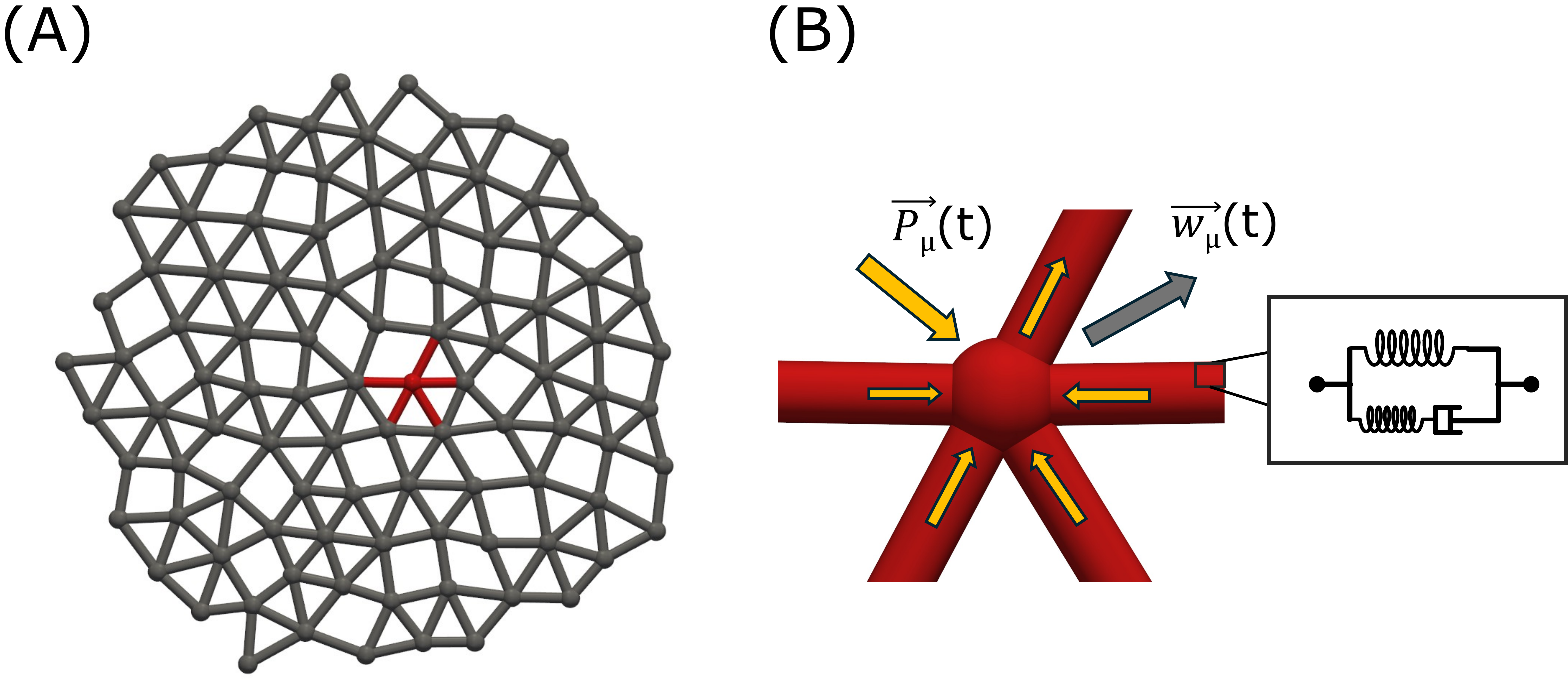}
    \caption{\textbf{Viscoelastic truss network model.}
\textbf{(A)} Example of a disordered two-dimensional truss network composed of viscoelastic rods connected by pin joints. One interior joint and its incident edges are highlighted in red to indicate the local neighborhood on which forces and displacements are defined.
\textbf{(B)} Zoomed-in view of the highlighted joint showing the applied force $\vec{P}_{\mu}(t)$ on the joint, internal stress in adjoining rods, and the resulting joint velocity $\vec{w}_{\mu}(t)$. Each infinitesimal element within the rod behaves as a standard linear solid (SLS), modeled as a spring in parallel with a Maxwell branch  (spring--dashpot).}
    \label{fig:schematic}
\end{figure}

Basically, the network Laplacian approach treats each rod in the truss structure (Fig. \ref{fig:schematic}) as a one-dimensional continuum whose axial stress and velocity fields can be solved exactly in Fourier space in time. Rather than discretizing rods into finite elements, the method keeps their full continuum dynamics and enforces two physical constraints at every joint: (i) the axial velocity of each rod at the joint must equal the projection of the joint velocity along that rod, and (ii) the vector sum of all axial forces along all rods connected to a joint must equal the applied force on that joint. Combining the closed-form rod solutions with these joint-level constraints yields a global linear relation between the vector of joint displacements and the vector of joint forces. The matrix that encodes this mapping is the network Laplacian. In the elastic case this Laplacian depends only on connectivity and material impedance; in the viscoelastic generalization it becomes frequency-dependent and complex-valued, incorporating energy dissipation along with the storage of elastic energy. The central advantage of this framework is that the size of the problem scales with the number of joints rather than with the number of rod discretization points, allowing accurate and efficient computation of dynamics even for large, disordered networks.\\

Following the construction of \cite{fancher2025analysis}, we have the following equations of motion for stress wave propagation in a single rod
\begin{subequations}
\begin{equation}
    \frac{\partial v}{\partial z}-\frac{\partial\epsilon}{\partial t} = 0,
    \label{matcon}
\end{equation}
\begin{equation}
    \frac{\partial\sigma}{\partial z}-\rho\frac{\partial v}{\partial t} = 0,
    \label{momcon}
\end{equation}
\label{coneqs}
\end{subequations}
where $v(z,t)$, $\epsilon(z,t)$, $\sigma(z,t)$ are uniaxial velocity, strain and stress field respectively, material density is denoted as $\rho$, and $z$ is a coordinate measured from one end of the rod. Here, we assume that the rods are viscoelastic, specifically, we consider the Boltzmann superposition principle for linear viscoelasticity \cite{bland1960theory, hilton2017elastic, boltzmann1878theorie} given by the relation
\begin{equation}
    \epsilon\left(t\right) = \int_{0}^{t}d\tau\>c\left(t-\tau\right)\frac{d\sigma\left(\tau\right)}{d\tau},
    \label{Boltzdef}
\end{equation}
\noindent where $\epsilon$ is the strain of the material, $\sigma$ is the stress, and $c$ is the creep function. The time $t=0$ is chosen such that $\sigma(t)=0$ for all $t<0$. This formulation can be rewritten by additionally enforcing that $c(t)=0$ for all $t<0$ as well. These null regions of $\sigma(t)$ and $c(t)$ allow the bounds of integration to be freely expanded into $[-\infty , \infty]$.
This then allows us to denote the Fourier transform of $\epsilon$ as $\tilde{\epsilon}\left(\omega\right) = 2\pi i\omega\tilde{c}\left(\omega\right)\tilde{\sigma}\left(\omega\right)$.\\

We then solve the wave equation and get solutions to velocity field ($\tilde{v})$ and internal stress ($\tilde{\sigma}$). For purposes of evaluating the dynamics of a rod of length $L$, the most useful forms of solutions to $\tilde{v}$ and $\tilde{\sigma}$ are given by

\begin{subequations}
\begin{equation}
    \tilde{v}\left(z,\omega\right) = \frac{\tilde{v}\left(0\right)\sinh\left(\left(L-z\right)\xi\right)+ \allowbreak \tilde{v}\left(L\right)\sinh\left(z\xi\right)}{\sinh\left(L\xi\right)},
    \label{velbound_vel}
\end{equation}
\begin{equation}
    \tilde{\sigma}\left(z,\omega\right) = -\frac{i\omega\rho}{\xi}\frac{\tilde{v}\left(0\right)\cosh\left(\left(L-z\right)\xi\right)- \tilde{v}\left(L\right)\cosh\left(z\xi\right)}{\sinh\left(L\xi\right)}.
    \label{velbound_sig}
\end{equation}
\label{velbound}
\end{subequations}
$\xi(\omega)$ is the dispersion function defined as

\begin{equation}
    \xi\left(\omega\right) \equiv \sqrt{2\pi\left(i\omega\right)^{3}\rho\tilde{c}\left(\omega\right)}.
    \label{xidef}
\end{equation}
\noindent

The particular form of solution in Eq. \ref{velbound} can be used to discretely connect dynamics of rods to that of the network.

\begin{subequations}
\begin{equation}
    \hat{e}_{\mu\nu}^{\text{T}}\vec{w}_{\mu}\left(t\right) = v_{\mu\nu}\left(0,t\right),
    \label{wvelrel}
\end{equation}
\begin{equation}
\vec{P}_{\mu}\left(t\right)+\sum_{\nu\in\mathcal{N}_{\mu}}\hat{e}_{\mu\nu}A_{\mu\nu}\sigma_{\mu\nu}\left(0,t\right) = \vec{0},
    \label{Psigrel}    
\end{equation}
\label{trusscons}
\end{subequations}
\noindent where $\vec{w}_{\mu}$ is the velocity of joint $\mu$, $\vec{P}_{\mu}$ is the externally applied force at joint $\mu$, $\hat{e}_{\mu\nu}$ is the unit vector pointing from joint $\mu$ to joint $\nu$, and $A_{\mu\nu}$ is the cross sectional area of the rod connecting joints $\mu$ and $\nu$. $v_{\mu\nu}(z,t)$ is the velocity field of rod $\mu\nu$ with the $z=0$ end of the rod being at joint $\mu$. 

We define multi-vectors $\mathbf{\vec{P}}$ and $\mathbf{\vec{W}}$ of
N components each, where N is the number of joints in the network. Their $\mu^{th}$ components are the Fourier transformed quantities, external force at  $\mu^{th}$ joint, $\vec{\tilde{P}}_{\mu}$ and joint velocity, $\vec{\tilde{w}}_{\mu}$  respectively. By combining Eq. \ref{velbound_sig} with Eq. \ref{trusscons} we can show that the joint force and velocity multivectors in Fourier space $\mathbf{\vec{P}}$ and $\mathbf{\vec{W}}$ are related via

\begin{equation}
    \mathbf{\vec{P}} = \frac{1}{i\omega}\mathbf{\overset\Leftrightarrow{D}}\mathbf{\vec{W}} \quad\quad \implies \quad\quad \mathbf{\vec{P}} = \mathbf{\overset\Leftrightarrow{D}}\mathbf{\vec{U}},
    \label{multimatrel}
\end{equation}

\noindent{where $\mathbf{\tilde{U}}$ is the joint displacement multivector of components N, where each component $\mu$ is the Fourier transformed displacement vector, $\vec{\tilde{u}}_{\mu}$. $\mathbf{\overset\Leftrightarrow{D}}$ is the computed network Laplacian (See details in Supplementary Information Section I).


\subsection{Incorporating standard linear solid model}
\label{sls}
The dispersion function, $\xi(\omega)$ is of critical importance to the problem of dissipation. Here we will consider a standard linear solid (SLS) and obtain an explicit form dispersion function. The creep function for such a SLS is given by
\begin{equation}
    c\left(t\right) = \frac{1}{E}\left(1-\left(1-\frac{\tau_{\epsilon}}{\tau_{\sigma}}\right)e^{-t/\tau_{\sigma}}\right)\Theta\left(t\right),
    \label{SLScreep}
\end{equation}
$\tau_e$ and $\tau_{\sigma}$ are two relaxation time constants, $E$ is the Young's modulus. The dispersion function can then be expressed as 
\begin{equation}
    \xi\left(\omega\right) = i\omega\sqrt{\frac{\rho}{E}\frac{1+i\omega\tau_{\epsilon}}{1+i\omega\tau_{\sigma}}}.
    \label{SLSxi}
\end{equation}
Analysis of $\xi(\omega)$ at real driving frequencies shows that
$|\xi_i| > |\xi_r|$ across the full frequency range, so physically
realizable operating points lie within the wedge $|\xi_r| < |\xi_i|$
in the complex-$\xi$ plane (See Supplementary Information, Fig.~S1).

\subsection{Rate of mechanical energy dissipation} 

The energy dissipated in each rod (i,j) can be computed analytically using the solutions to the joint velocities $\mathbf{\vec{W}}$ obtained by solving Eq. \ref{multimatrel}. Then, for each rod connecting joints \(i\) and \(j\), we project the local velocity field along the unit vector \(\hat{\mathbf{e}}_{ij}\) to obtain the scalar velocities at the ends:
\begin{equation}
\tilde{v}(0,\omega)= \mathbf{\tilde{w}}_i \cdot \hat{\mathbf{e}}_{ij}, \quad
\tilde{v}(L, \omega) = \mathbf{\tilde{w}}_j \cdot \hat{\mathbf{e}}_{ij}.
\label{eq:vbc_proj}
\end{equation}

The stress-power density in the material is
\begin{equation}
    \mathcal{P}\left(z,t\right) = \sigma\left(z,t\right)\frac{\partial\epsilon\left(z,t\right)}{\partial t},
    \label{qdensitydef}
\end{equation}
For viscoelastic materials, this stress power density includes changes in stored elastic energy and the mechanical energy dissipated as heat. We now restrict our consideration to only periodic dynamics with well defined period $T$, with stress and strain represented as 
\begin{subequations}
    \begin{equation}
        \sigma\left(z,t\right) = \sum_{n=-\infty}^{\infty}\tilde{\sigma}_{n}\left(z\right)e^{2\pi int/T},
        \label{sigperiodic}
    \end{equation}
    \begin{equation}
        \epsilon\left(z,t\right) = \sum_{m=-\infty}^{\infty}\tilde{\epsilon}_{m}\left(z\right)e^{2\pi imt/T},
        \label{sigperiodic}
    \end{equation}
    \label{sigepperiodic}
\end{subequations}

\noindent where $\tilde{\sigma}_{n}$ and $\tilde{\epsilon}_{m}$ are Fourier coefficients that represent amplitudes and phase of frequency components making up the periodic stress and strain waveforms respectively. Because of this periodicity, the sum of elastic and kinetic energy remains constant over the time period $T$. Any cumulated change in mechanical energy over time $T$ represents energy dissipated as heat which is equal to the power expended by force acting at the boundary (See Supplementary Information, Fig.~S2). In non time periodic systems, the rate of energy dissipation has to be computed as $\frac{1}{2}\eta \dot{\epsilon}^2$, where $\eta$ is the viscosity in the Maxwell element. Thus, the average rate of energy dissipated per unit length, $\bar{q}(z)$ is
\begin{equation}
    \bar{q}(z) = -\frac{A_{ij}}{T}\int_{0}^{T}dt\>\mathcal{P}\left(z,t\right)
\end{equation}
The averaged rate of energy dissipated by a rod (i,j) is therefore
\begin{equation}
    Q_{ij}= \int_{0}^{L}dz \bar{q}(z)
\end{equation}
On substitution,
\begin{equation}
    Q_{ij}= \frac{2\left(2\pi\right)^{3}A_{ij}}{T^{2}}\sum_{n=1}^{\infty}n^{2}\text{Re}\left(\tilde{c}_{n}\right)\int_{0}^{L}dz\>\left|\tilde{\sigma}_{n}\left(z\right)\right|^{2}
    \label{Qepred}
\end{equation}
Substituting the expression for stress from Eq. \ref{velbound_sig}, we can simplify further and obtain a closed form expression (See Supplementary Information Section II for details). For a single rod driven at one end with sinusoidal displacement (Fig.~\ref{fig:combined_two_panel}A), $u(0)=0.001\sin t$, we solved for $\bar{q}(z)$ analytically. $\bar{q}(z)$ has a non-oscillatory envelope proportional to $\cosh(2\xi_r (L-z))$ and an oscillatory modulation $\cos(2\xi_i (L-z))$ as a function of distance along the rod $z$ (Fig.~\ref{fig:combined_two_panel}B). For $|\xi_r|>0$, the envelope decays exponentially near $z=0$, defining the attenuation length
\begin{equation}
\ell = \frac{1}{2|\xi_r|}.
\label{eq:ell}
\end{equation}
The oscillatory term has a spatial period $\lambda$ which is given by
\begin{equation}
\lambda = \frac{\pi}{|\xi_i|}.
\end{equation}
Fig.~\ref{fig:combined_two_panel}B shows $\bar{q}(z)$ for representative values of the complex wavenumber $\xi = \xi_r + i\xi_i$. The range of allowed values for a linear viscoelastic material are $\xi_i >0$, $\xi_r < 0$ obtained by requiring $\bar{q}(z)$ to be positive. As noted before $|\xi_i |> |\xi_r|$ (See Section \ref{sls}). If $\xi_r=0$, the energy dissipated is 0, which corresponds to the case of purely elastic systems. The peak amplitude of $\bar{q}(z)$ depends strongly on $\xi_i$ and is largest near the global resonant modes of lossless systems $\xi_i = n\pi/L$ (See Supplementary Information, Section III A).

The total energy dissipated by the rod, $Q_{01}$, increases approximately 
linearly with $|\xi_r|$ at small $|\xi_r|$, reaches a maximum, and decreases at large $|\xi_r|$ (Fig.~\ref{fig:combined_two_panel}C).  The curve at $\xi_i = 3.15$ in Fig.~\ref{fig:combined_two_panel}C is near the $n=10$ resonance of lossless system
(at $\xi_i = \pi$ for $L=10$) and accordingly exhibits the largest total energy dissipated for low values of $\xi_r$. We note that this large $Q$ reflects the resonant amplification of internal stress and strain fields within the rod near resonance. Although the boundary displacement (u(0)) is held fixed, the internal response diverges at resonant wavenumbers, $\xi_i = n\pi/L$. This large internal displacement leads to high energy dissipation for finite $\xi_r$ values. Taking the limit $\xi_r \to 0$ at fixed $\xi_i = n\pi/L$ gives $Q_{01} \sim 1/|\xi_r|$ (See Supplementary Information, Section III A). The divergence at $\xi_r = 0$ is unphysical because dissipation requires $\xi_r > 0$. At $\xi_r = 0$ there is no dissipation mechanism and $Q_{01}=0$. Practically this divergence is regulated by finite $\xi_r$ and by the breakdown of the small-amplitude assumption in materials near resonance.

\begin{figure}
        \centering
        \includegraphics[width=0.9\linewidth]{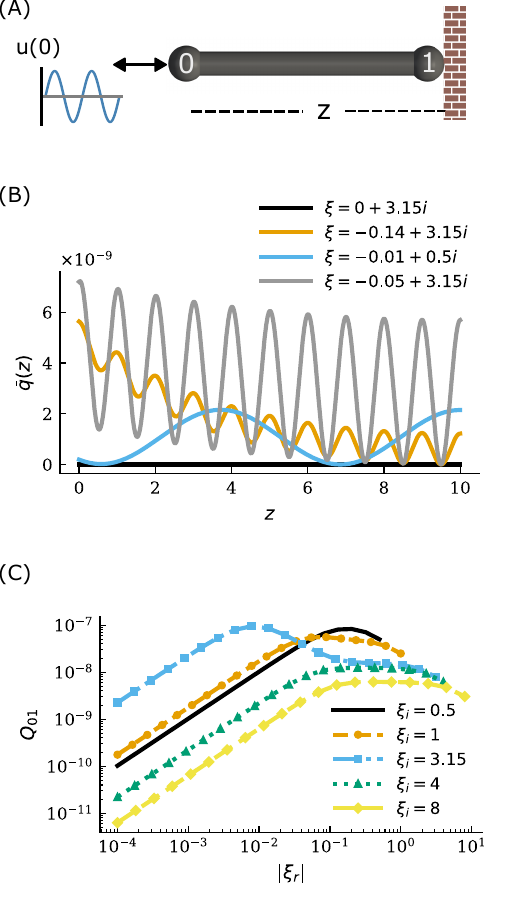}
        \caption{\textbf{Dissipation in 1D bulk material} \textbf{(A)} A viscoelastic rod of length $L$, clamped at one end and driven harmonically at the other. The coordinate $z$ denotes distance from the driven end. \textbf{(B)} Spatial profile of the energy dissipated per unit length per unit time, $\bar{q} (z)$, for several values of the complex wavenumber $\xi = \xi_r + i\xi_i$. \textbf{(C)} Total rate of energy dissipated by the rod as a function of the real part of complex wave number. $L=10$, $A=0.1$, $\rho=1$, $u(0) = 0.001 \sin t$.}
        \label{fig:combined_two_panel}
\end{figure}

\subsection{Benchmarking numerical implementation}

\begin{figure*}
    \centering
    \includegraphics[width=\linewidth]{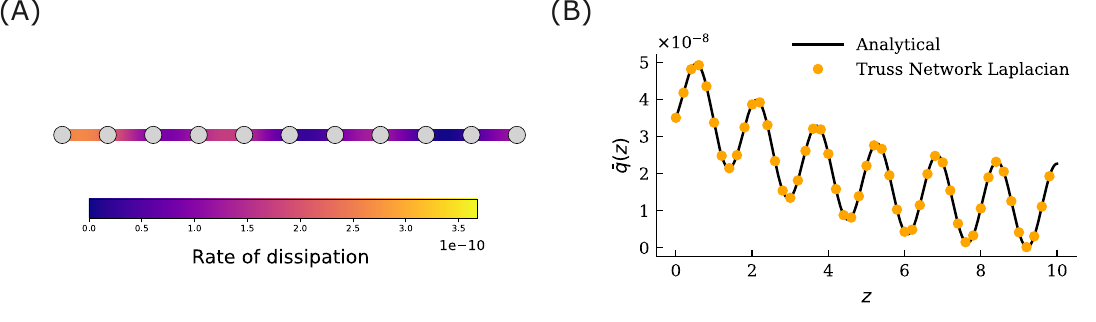}
    \caption{\textbf{Benchmarking: energy dissipation in a uniform 1D network.} \textbf{(A)} Rate of energy dissipation per segment in a one-dimensional network of 10 rods each of equal length ($L=1$) and cross sectional area ($A=1$), driven harmonically at the leftmost joint and fixed at the rightmost joint. Each rod was discretized into $N_{seg}=500$. \textbf{(B)} Comparison of rate of energy dissipation per unit length, $\bar{q}(z)$ computed by the graph Laplacian based spectral framework against the analytical solution for a single rod, confirming exact agreement. $\xi = -0.1+2i$, $\rho=1$, $\omega=1$, $u(0,t) = 0.001 \sin t$, $u(10, t)=0$. }
\label{fig:one_dimensional_benchmarking_and_area_analysis}
\end{figure*}

We first verify the accuracy of the graph Laplacian-based spectral framework for truss networks by comparing numerical results against the analytical solution for a single rod. To this end, we model a rod of total length $10$ as a one-dimensional network of $10$ rods of length $1$ each. The driven joint was subjected to a sinusoidal displacement, $u(0,t) = 0.001 \sin t$, the rightmost joint was held fixed, $u(10, t)=0$.

The energy dissipation rate in the one-dimensional network obtained from computational solver oscillates with decaying amplitude as a function of distance from the driven joint (Fig.~\ref{fig:one_dimensional_benchmarking_and_area_analysis}A) and matches with analytical expression exactly (Fig.~\ref{fig:one_dimensional_benchmarking_and_area_analysis}B), demonstrating the accuracy of numerical implementation (See Supplementary Information, Fig.~S3 for a different parameter set).

\section{Cross-sectional area as a design variable for energy dissipation}
\label{sec:optimization}
We next use the graph Laplacian-based spectral framework developed in Sec.~\ref{truss_formalism} to understand how spatially redistributing mass with cross-sectional area as the design variable affects the attenuation length and total energy dissipated ($Q_\mathrm{total}$). We focus on a base material with $\xi_i$ close to a global resonance mode, specifically $\xi_i=3.15$ for the rest of the paper, so that total energy dissipated is high.

\begin{figure*}
    \centering
    \includegraphics[width=\linewidth]{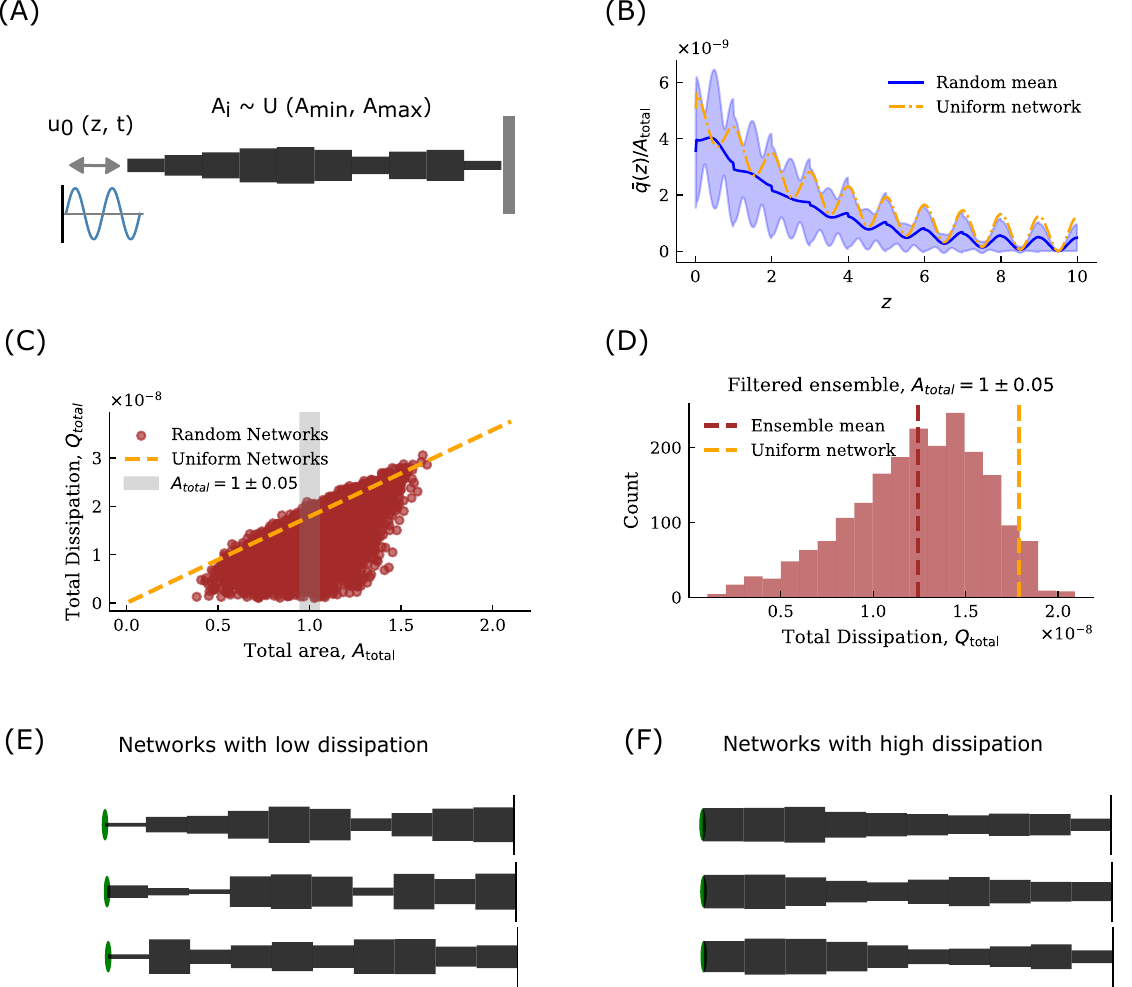}
    \caption{\textbf{Total energy dissipation in one-dimensional networks with random cross-sectional areas.} \textbf{(A)} A one-dimensional network of 10 rods of equal length (L), with cross-sectional areas drawn independently from a uniform distribution $U(A_\mathrm{min}, A_\mathrm{max})$. The leftmost joint is driven harmonically with amplitude $\tilde{u}(0)$ and frequency $\omega$ as $u(0,t)=\tilde{u}(0) \sin \omega t$ and the rightmost joint is fixed. \textbf{(B)} Rate of energy dissipation per unit length per unit time , $\bar{q} (z)$, normalized by total network area, $A_\mathrm{total} = \sum_{\langle i, j \rangle} A_{ij}$, as a function of distance from the source (driven joint), $z$. The solid blue line is the mean over all realizations; the orange dash dotted line is the result for a uniform network with all rod areas equal. Number of random realizations is 10000.\textbf{(C)} Total energy dissipated $Q_\mathrm{total}$ plotted against total network area $A_\mathrm{total}$. The orange dashed line shows $Q_\mathrm{total}$ for uniform networks of varying area, included for reference.   \textbf{(D)} Distribution of $Q_\mathrm{total}$ for the subset of realizations with $A_\mathrm{total} = 1 \pm 0.05$. The distribution is negatively skewed: most realizations dissipate less energy than the uniform network (orange dashed line), with only rare configurations exceeding it. Red dashed line is the ensemble mean. \textbf{(E, F)} Representative network architectures with low ($Q_\mathrm{total} < 5\times 10^{-10}$) and high ($Q_\mathrm{total} > 2.0 \times 10^{-8}$) dissipation. Rod thickness is proportional to the square root of cross-sectional area. Low-dissipation networks have thin rods near the source; high-dissipation networks have thick rods near the source followed by progressively thinner rods. $L = 1$. $A_{min}=0.0012$. $A_{max}=0.2$. $\tilde{u}(0) = 0.001$. $\omega=1$. $\rho=1$ and $\xi = -0.14 + 3.15i$ for all rods in the networks.}
    \label{fig:total_energy_1d_network}
\end{figure*}

\subsection{Random rod cross-sectional areas}
We examine energy dissipated in a one-dimensional network in which rod cross-sectional areas are assigned randomly (Fig.~\ref{fig:total_energy_1d_network}A), at the wavenumber $\xi_i=3.15$. The areas are drawn from a uniform distribution bounded between a minimum ($A_\mathrm{min}$) and maximum ($A_\mathrm{max}$) value. We also computed the dissipation for a uniform network whose area equals the average of these extrema.

We normalized the energy dissipated by the total area of the network, $A_{total} = \sum_{\langle i, j \rangle} A_{ij}$, the summation is over all unique joint pairs $(i,j)$ (Fig.~\ref{fig:total_energy_1d_network}B). Interestingly, despite the randomness in cross-sectional areas, the attenuation length and wavelength remain largely unchanged compared to the uniform network, as shown by visual inspection of the ensemble mean of the dissipation profile compared to the uniform network (Fig.~\ref{fig:total_energy_1d_network}B).

We then computed the the total energy dissipated per unit time by the network (sum of energy dissipated by all rods in networks) (Fig.~\ref{fig:total_energy_1d_network}C) and plotted against $A_{total}$. In addition the energy dissipated for uniform networks of different $A_{total}$ is included. For uniform networks, $Q_{total}$
scales linearly with $A_{total}$, a consequence of the displacement boundary condition that we impose. For example, the energy dissipated by two rods under identical displacement boundary conditions equals that of a single rod with an area equal to their sum. Interestingly, for a given $A_{total}$, we see $Q_\mathrm{total}$ changes with the random realization, and there are several realizations with energy dissipated higher than uniform networks, thus indicating that areas of cross section of rods can be used as a design variable to change energy dissipation. 

To probe this further, we filtered out all realizations with $A_{total} = 1\pm 0.05$ and obtained a distribution of total energy dissipated $Q_\mathrm{total}$ (Fig.~\ref{fig:total_energy_1d_network}D). Clearly the ensemble mean of energy dissipated is lower than that of uniform networks. This suggests that if one were to select a particular distribution of cross sections at random, the resulting structure is more likely to dissipate less energy than a uniform network. 

The networks that dissipate less energy, tend to have very thin rods closer to the source (Fig.~\ref{fig:total_energy_1d_network}E). This leads to localized oscillations that do not propagate to the rest of network. In other words, it requires less energy input for the source to achieve desired displacement boundary conditions at the input joint if the first edge is thin. Low energy input to the network in turn leads to lower dissipation. On the other hand, networks that dissipate more energy tend to have a gradient architecture with thick rods near the source followed by thinner and thinner rods (Fig.~\ref{fig:total_energy_1d_network}F) as we move away from the source. This is a consequence of the global resonance mode at this driving frequency ($\xi_i=3.15$) where the stress wave amplitude is $\lambda\approx 1$, equal to the edge length.

Gradient architecture is also observed for other global resonant frequencies such as $\xi_i=n \pi/L_{net}$, where $L_{net}$ is the length of our network along the direction of excitation (See Fig.~S5 for $\xi_i = 3\pi /L_{net}$). Outside these special frequencies, optimal architectures show no discernible spatial pattern (Supplementary Information, Fig. S4).
\begin{figure*}
    \centering
    \includegraphics[width=0.9\linewidth]{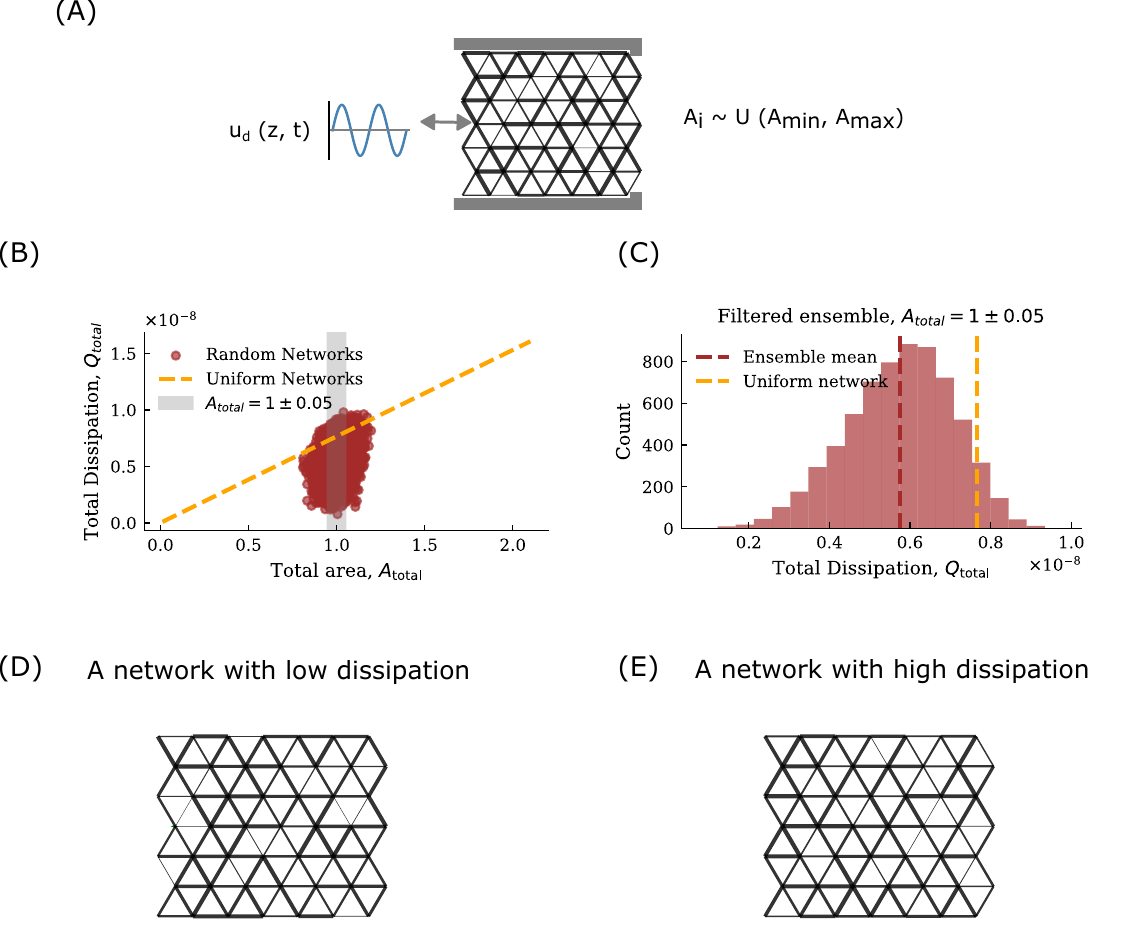}
    \caption{\textbf{Total energy dissipation in two-dimensional networks with random cross-sectional areas.} \textbf{(A)} A two-dimensional triangular truss network with cross-sectional areas drawn independently from a uniform distribution $U(A_\mathrm{min}, A_\mathrm{max})$. All rods have equal length $L = 1$ and the same material properties. The arrow indicates the driven joint, excited harmonically along the $x$-axis with amplitude, $\tilde{u}_d$ and frequency, $\omega$. Roller boundary conditions applied for top and bottom surface as indicated. Rightmost top and bottom nodes are clamped. \textbf{(B)} Total energy dissipated ($Q_\mathrm{total}$) for random networks plotted against total network area (($A_\mathrm{total}$). The orange line shows results for uniform networks of varying area, included for reference. Number of random realizations is 10000. \textbf{(C)} Distribution of $Q_\mathrm{total}$ for the subset of realizations with $A_\mathrm{total} = 1 \pm 0.05$. As in the one-dimensional case, most realizations dissipate less than the uniform network baseline. \textbf{(D, E)} Representative network architectures with low and high dissipation. Rod thickness is proportional to the square root of cross-sectional area. $\tilde{u}_d = 0.001$, $\omega=1$,  $\rho=1$ and $\xi = -0.14 + 3.15i$ for all rods in the networks. }
    \label{2d_network_analysis_area}
\end{figure*}

We next extended the one-dimensional analysis to a two-dimensional triangular network with random rod cross sectional areas (Fig.~\ref{2d_network_analysis_area}A). A harmonic displacement boundary condition was applied at a single joint located at the leftmost, vertically central position of the network (which we refer as driven or source joint) with the displacement imposed along the $x$-direction. Distribution of total energy dissipated has a similar behavior like in 1D random networks, Fig.~\ref{2d_network_analysis_area}B, and networks with low dissipation tend to have thin rods near source and networks with high dissipation tend to have thick rods near the source. As before, this source weighted architecture is less evident in general cases of $\xi_i$ (See Supplementary Information, Fig.~S6).

 Overall, this indicates that the qualitative features identified in one-dimensional systems persist in two-dimensional networks, suggesting that the observed trends are robust to changes in network dimensionality.

\subsection{Optimization of area of cross section for maximizing energy dissipated}
Next, we optimized 2D network architectures to maximize the total viscoelastic energy dissipation under a fixed material cost constraint by treating rod cross-sectional areas as continuous design variables.  We consider the material cost as
\begin{equation}
    C(\boldsymbol{A}) = \sum_{\langle i, j \rangle} A_{ij}^{\gamma},
\end{equation}
We enforce a fixed total cost $C(\boldsymbol{A}) = C_{\rm target}$ through a Lagrange multiplier correction combined with a projection of the rod areas onto the constraint manifold. Here $\textbf{A}$ is the matrix of rod cross sectional areas.

The optimization objective function is the total energy dissipated by the network per unit time.
\begin{equation}
    Q_\mathrm{total} = \sum_{\langle i,j \rangle} Q_{ij}(\tilde{\mathbf{u}}, \textbf{A}),
\end{equation}
where $Q_{ij}$ is computed from the viscoelastic stress and strain amplitudes along each rod using Eq. \ref{Qepred}. We then minimize the loss functional
\begin{equation}
    \mathcal{L}(\boldsymbol{A}) = -Q_\mathrm{total}(\boldsymbol{A})
    + \lambda\,\big(C(\boldsymbol{A}) - C_{\rm target}\big),
\end{equation}
The Lagrange multiplier was computed for each step as in \cite{chang2019microvascular}
\begin{equation}
    \lambda
    =
    -\frac{
        \displaystyle \sum_{\langle i, j \rangle} 
        A_{ij}^{\gamma-1} \, \frac{\partial Q_\mathrm{total}}{\partial A_{ij}
    }}{
        \displaystyle \gamma 
        \sum_{\langle i, j \rangle} A_{ij}^{2(\gamma-1)}
    }.
    \label{eq:lambda_expression}
\end{equation}
In this work we take $\gamma =1$. 

A gradient step $A_{ij}\leftarrow A_{ij}-\alpha\partial\mathcal{L}/\partial A_{ij}$ is followed by clipping areas back into the admissible interval and projected to constrain the manifold to satisfy the material-cost constraint. The learning rate $\alpha$ is reduced by a constant factor, $l_f$, whenever the objective fails to improve by more than a relative tolerance $\epsilon_\text{rel}$ for $N_\text{plateau}$ consecutive iterations, and the optimization terminates when $\alpha$ reaches its minimum value $\alpha_\text{min}$ and loss function converges. The area vector, \textbf{A}, achieving the maximum $Q_\mathrm{total}$ over the entire run is recorded as the optimal solution. 

Figure~\ref{fig:one_optimization_run_for_illustration} shows a representative optimization trajectory starting from a random distribution of cross-sectional areas. As the optimization proceeds, the total dissipated energy increases monotonically through a systematic redistribution of material toward more dissipative configurations at fixed total mass. The optimized architectures are source-weighted and reticulated (loopy) as evident from visual inspection: mass is concentrated near the driven joint, and cycles are present across the network. Both features have a straightforward physical interpretation. The loopy connectivity maintains structural rigidity, allowing the network as a whole to provide sufficient mechanical impedance at the source for energy to be absorbed under prescribed displacement boundary conditions. The concentration of mass near the source enhances dissipation directly, since at a fixed network configuration the energy dissipated by an edge increases linearly with its cross-sectional area (Eq.~\ref{Qepred}). 

These optimized architectures qualitatively resemble flow networks optimized for transport efficiency~\cite{durand2007structure,bohn2007structure, banavar2000topology,katifori2010damage}: both develop gradient structures that taper away from the source. However, there are several differences between these optimization problems worth noting. First, optimal flow networks involve differences in pressure values at nodes, which are purely scalar instead of velocities in mechanical systems which are vector quantities. Optimal flow networks analogous to our problem are tree-like~\cite{durand2007structure,bohn2007structure, banavar2000topology}, whereas our optimal mechanical networks remain loopy, a consequence of the rigidity requirement. Finally, the two optimization problems appear to invert one another: we \emph{maximize} the dissipated power by varying cross-sectional areas, whereas flow networks \emph{minimize} dissipated power by varying edge conductivities. This apparent contradiction is resolved by noting the differences in boundary conditions: prior flow-network studies fix the input current, while we fix the input displacement (the analogue of fixing pressure in fluid flow networks or voltage in resistor networks \cite{firestone1933new, ortiz2025unified}).
\begin{figure}
    \centering
    \includegraphics[width=\linewidth]{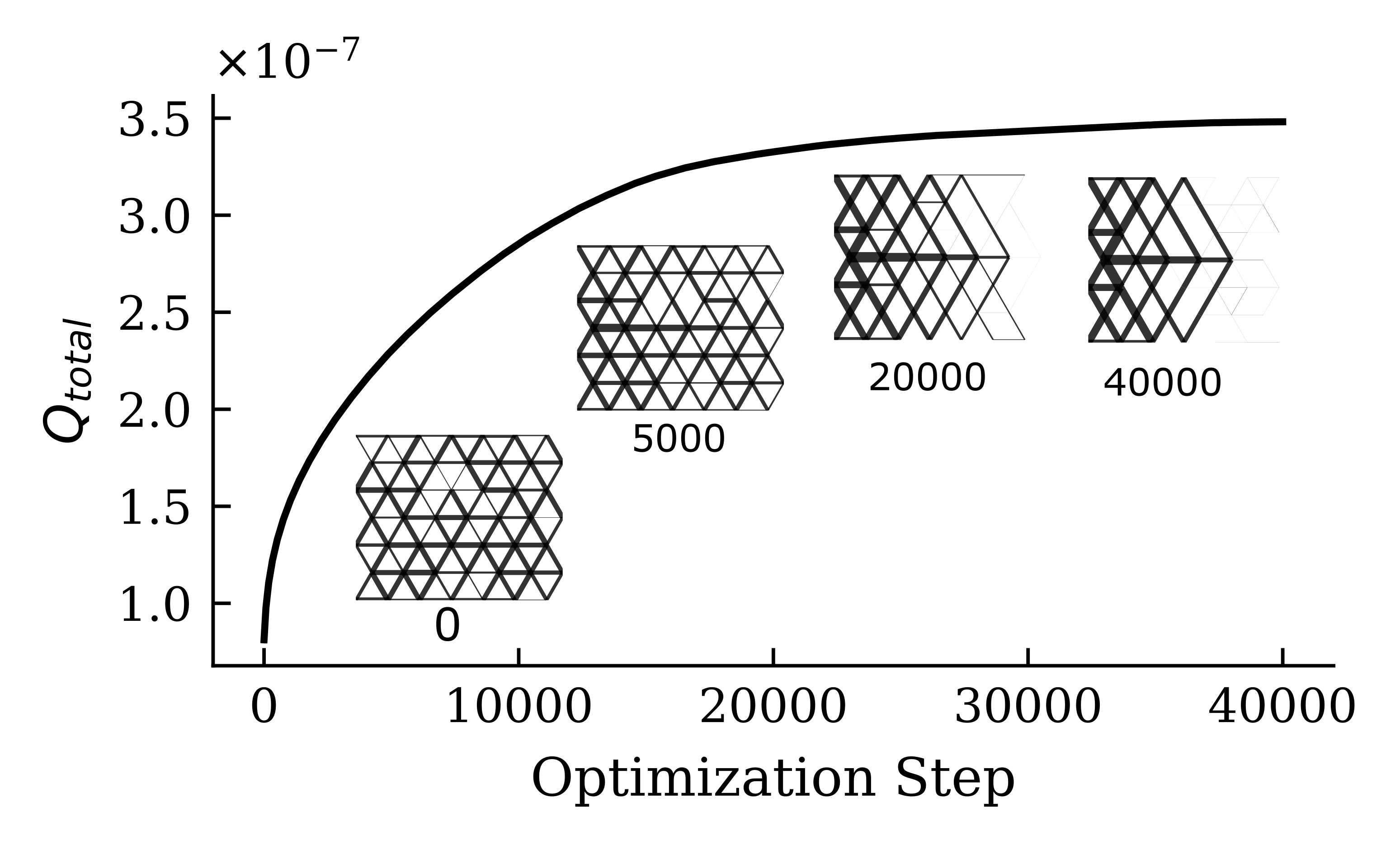}
    \caption{\textbf{Illustrative optimization trajectory for a two-dimensional viscoelastic network.} Evolution of a single gradient-based optimization run maximizing the total energy dissipated under a fixed material cost constraint. The solid black line is rate of total energy dissipated by the network as a function of optimization step. Insets depict the network architecture at selected steps (labeled), with rod thickness proportional to square root of cross-sectional area (divided by square root of cross-sectional area of thickest rod in the network). Starting from a random initial configuration, the optimization monotonically increases dissipation and converges to a source-weighted, loopy architecture. Network parameters: $\xi=-0.01+3.15i$, $\rho=1$, $L=1$ for all rods. Same boundary conditions as Fig.~\ref{2d_network_analysis_area}. Optimization parameters: $\alpha=\alpha_{min}=5$, $\epsilon_{rel}= 10^{-6}$, $N_{plateau}=200$, $C_{target}=1$.}
    \label{fig:one_optimization_run_for_illustration}
\end{figure}

\subsection{Optimal network structure is sensitive to the base material}
\begin{figure*}
    \centering
    \includegraphics[width=0.75\linewidth]{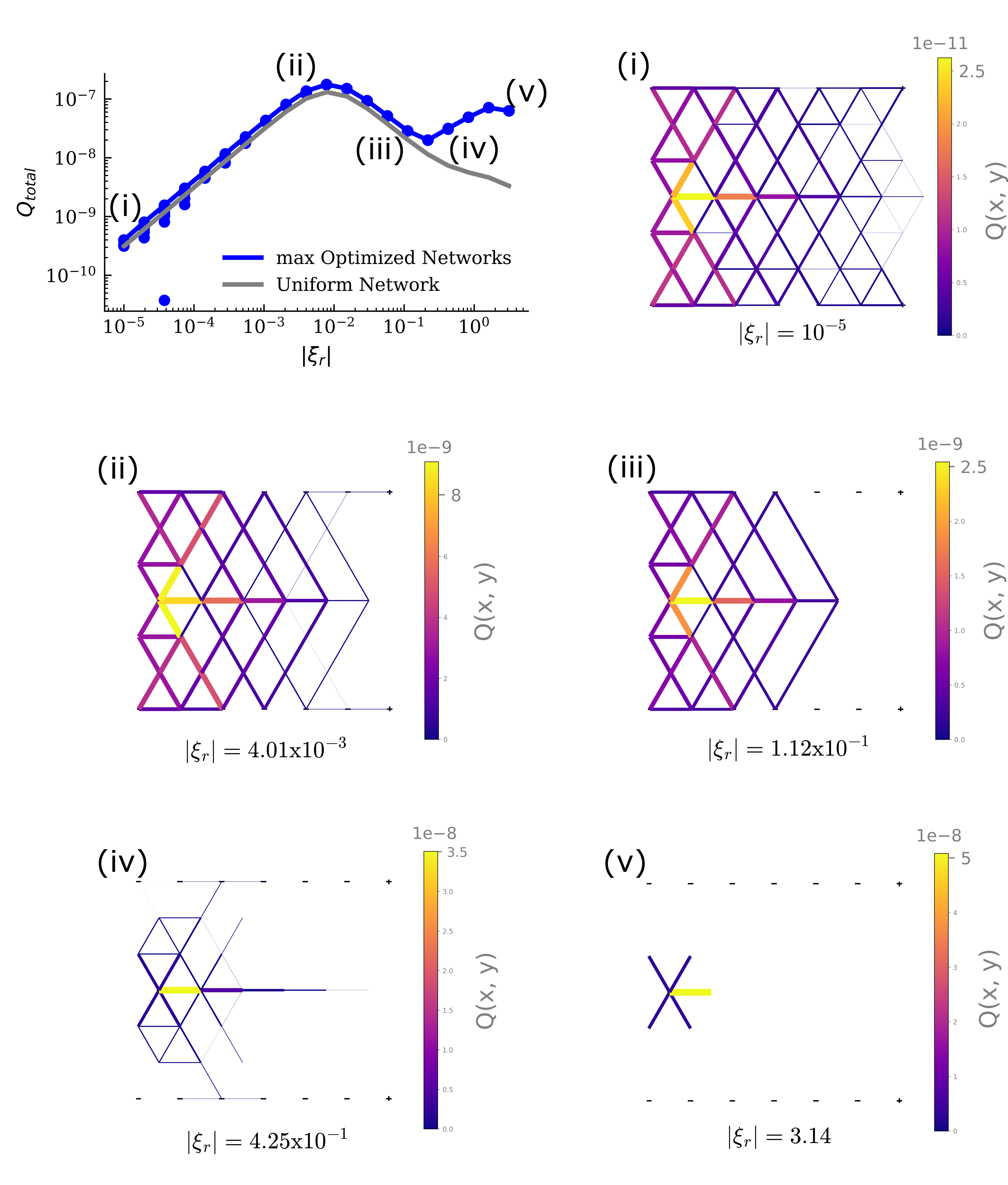}
    \caption{\textbf{Optimal network architecture depends on intrinsic material dissipation.} (Top left) Total energy dissipated per unit time, $Q_\mathrm{total}$, as a function of the real part of the complex wavenumber $\xi_r$, which controls the intrinsic attenuation length of the viscoelastic material. Points show results from optimization runs initialized from random area distributions; the solid line connects the maximum value among these runs. Figs i-v show representative optimized network architectures at selected values of $|\xi_r|$, with rod thickness proportional to the square root of cross-sectional area divided by the square root of area of the thickest rod in the network, and color indicating energy dissipated by rods. Symbol `$-$' indicates roller boundary condition along horizontal axis and `$+$' indicates fixed boundary condition. As $\xi_r$ increases, $Q_\mathrm{total}$ exhibits two maxima, accompanied by qualitative changes in optimal network architectures. Network parameters: $\xi_i=3.15$, $\rho=1$, $L=1$ for all rods in networks. Same boundary conditions as Fig.~\ref{2d_network_analysis_area}. Optimization parameters: $\alpha= 500$, $\alpha_{min}=0.05$, $l_f=0.1$, $N_{plateau}=200$, $\epsilon_{rel}=10^{-6}$, $C_{target}=1$, Number of samples for each $\xi_r$, $N_{trials}=10$. Colorbar range is set independently per panel.}
    \label{fig:optimize_sweep_xr}
\end{figure*}
We next investigated how the optimal network architecture depends on the intrinsic dissipation properties of the base material. To this end, we varied the real part of the complex wavenumber $\xi_r$ while keeping $\xi_i$ fixed, and computed the corresponding optimal structures (Fig.~\ref{fig:optimize_sweep_xr}).

For a uniform network, $Q_\mathrm{total}$, has a single global maximum at intermediate $\xi_r$. For small $\xi_r$ the material is weakly dissipative and behaves nearly elastically, while for large $\xi_r$ viscosity dominates and the material cannot store elastic energy efficiently. This is in agreement with analytical calculation in Fig.~\ref{fig:combined_two_panel} for a bulk material. Gradient-based optimization improves upon the uniform network across the entire range of $\xi_r$, and surprisingly reveals a second optimum at large $\xi_r$, a regime that is only an inflexion point for uniform networks. This second optimum corresponds to the case where the attenuation length $\ell = 1/(2|\xi_r|)$ is approximately equal to one edge length, so that the energy is almost entirely absorbed within rods immediately near the source joint. As $\xi_r$ is decreased further, the material cannot be further concentrated leading to a slight reduction in $Q_\mathrm{total}$. Hence this optimum arises because of the discrete lattice.

For all values of $\xi_r$, the optimal networks exhibit a gradient architecture in which rod thickness decreases gradually moving away from the excitation source (Fig.~\ref{fig:optimize_sweep_xr} i-v). Visual inspections of optimal structures indicate that the steepness of this gradient is correlated with $\xi_r$ of base material. This is consistent with the spatial profile of stress-wave amplitudes in a driven rod. For small $\xi_r$, the attenuation length is large and stress-wave amplitude is expected to be substantial throughout entire network, hence one would expect broadly distributed structures to maximize energy dissipation. As $\xi_r$ increases, the attenuation length decreases and the stress-waves are expected to penetrate less, thereby it is best to distribute the given mass within this penetration areas to obtain maximal dissipation. Near the second optimum, most of the injected energy is absorbed within the first few edges, and mass is concentrated on rods immediately near the source. For very large $\xi_r$ ($<|\xi_i|$), viscous effects dominate and energy decays so rapidly that material beyond edges immediately connected to source contributes negligibly to dissipation. 

We also note that as the attenuation length becomes small (as $\xi_r$ increases), the optimal architectures tend to have masses closer to source. We were curious to know whether in such cases, could it also be that the boundary conditions at the walls far away from source have minimal influence on the optimal architectures. To test that, we computed optimal architectures for a different boundary condition where nodes at the topmost and bottommost rows are fixed (See Supplementary Information, Fig.~S9). Indeed,  optimal architectures at small $\xi_r$ values are sensitive to the boundary conditions whereas at large $\xi_r$ values they are less sensitive.}

\begin{figure*}[t]
    \centering
    \includegraphics[width=\linewidth]{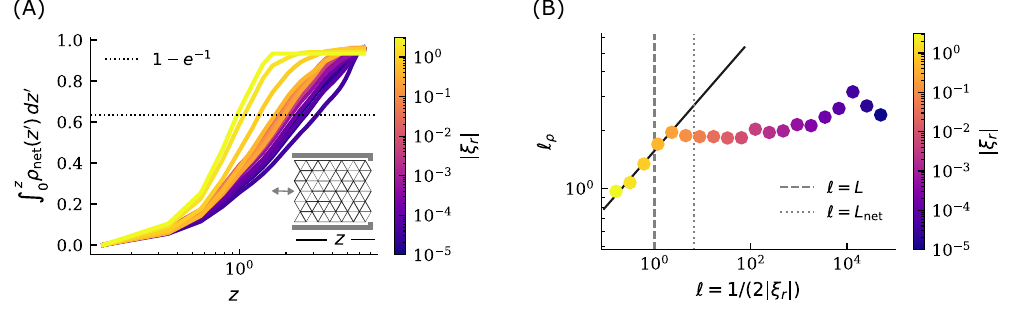}
    \caption{
    \textbf{Characterization of optimal networks.} \textbf{(A)} Cumulative mass $c(z) = \int_0^z \rho_{\rm net}(z')\mathrm{d}z'$. where $\rho_{\rm net}(z)$ is vertically projected linear mass density in the optimal networks (from analysis in Fig.~\ref{fig:optimize_sweep_xr}) as a function of horizontal position $z$. Each curve is the trial-averaged profile over $N_{\rm trials}=10$ independent optimizations; color encodes $|\xi_r|$ on a logarithmic scale (20 values spanning $|\xi_r| \in [10^{-5}, 3.14]$). The dotted horizontal line marks $1-e^{-1}\approx 0.632$; its intersection with each curve defines the characteristic length scale $\ell_\rho$, referred in this work as optimal density decay length. \textbf{(B)} Optimal density decay length, $\ell_\rho$ versus the attenuation length of base material $\ell = 1/(2|\xi_r|)$ for the same sweep. Solid black line guides the eye on scaling between $\ell_{\rho}$ and $\ell$. Dashed line marks edge length ($L$) and dotted line marks system size ($L_{net}$), where finite-size effects are expected to limit the scaling.} 
    \label{fig:optimal_network_characterization}
\end{figure*}

To quantify the spatial distribution of material in optimal networks and to check its correlation with $\xi_r$ of base material, we computed the mean mass density profile $\rho_\mathrm{net}(z)$ along vertical cross-sections of the network along $z$ - the horizontal coordinate along the network. To calculate $\rho_\mathrm{net}(z)$, each rod $(i,j)$ is discretized into $N_{\rm seg}=10$ equal segments along its length; the segment at horizontal position $z_{\rm mid}$ contributes mass $\rho A_{ij}\mathrm{d}s$, where $\mathrm{d}s = L_{ij}/N_{\rm seg}$ is the arc-length element. There are no vertical rods in the networks considered. Segment masses are accumulated into $3N_{\rm seg}$ uniform bins spanning the horizontal extent of the network and divided by the bin width, yielding $\rho_{\rm net}(z)$ with units of mass per unit length. The reported profile is the mean over $N_{\rm trials}$ trials (See Supplementary Information, Fig.~S7).
Then the cumulative mass profile $c(z) = \int_0^z \rho_{\rm net}(z')\mathrm{d}z'$ is computed via the trapezoidal rule on the binned data (Fig.~\ref{fig:optimal_network_characterization}A). Since the mass constraint fixes the total network mass to unity, $c(z=L_{net})$ approaches 1 (Fig.~\ref{fig:optimal_network_characterization}A). We define a length scale, $\ell_\rho$ as the position at which the cumulative mass reaches the fraction $1 - e^{-1} \approx 0.632$, i.e., $c(\ell_\rho) = 1 - e^{-1}$. In Fig.~\ref{fig:optimal_network_characterization}B  we compare the calculated $\ell_{\rho}$ to the theoretical penetration depth, the attenuation length $\ell$. We find that $\ell_{\rho}$ scales with $\ell$ demonstrating that it is the mechanical energy penetration that controls the optimal mass distribution. At $\ell=L_{net}$ when $\ell$ becomes the order of system size, the $\ell_{\rho}$ saturates. The $\ell_{\rho}$  vs $\ell$ scaling is robust to small and large perturbations in lattice positions (See Supplementary Information, Fig.~S10 and Fig.~S11).

Beyond the gradient architecture, optimal networks exhibit significant nearest-neighbor structural correlations. We quantified these using the two-point correlation computed over all unique nearest-neighbor edge pairs $A_{ij}$ and $A_{ik}$ (See Supplementary Information, Fig.~S8). The correlation $\langle S \rangle$ increases with $\xi_r$ up to the first maximum and then decreases, reflecting the growing spatial order imposed by the optimizer as energy becomes more localized. Taken together, these results establish that the optimal mass distribution follows the attenuation length scale of the base material under driving at certain resonant frequencies.


\section{Conclusion}
\label{sec:conclusion}
In this work, we extended the graph Laplacian-based spectral solver of~\cite{fancher2025analysis} to viscoelastic materials. The key ingredient is the Boltzmann formulation of linear viscoelasticity, which in Fourier space yields a linear stress-strain relation with a frequency-dependent complex elastic modulus. We derived this modulus explicitly for the standard linear solid model. We used this relationship to generalize the joint-level force balance and velocity compatibility conditions to the viscoelastic setting of \cite{fancher2025analysis}. The result is a frequency-dependent, complex-valued network Laplacian that relates temporally Fourier-transformed joint velocities to joint forces.

A key computational advantage of this framework is that the matrix size scales as $n = dN$, where $N$ is the number of joints and $d$ is the spatial dimension, rather than with the number of element-level discretization points as in conventional finite element methods. Moreover, direct solvers for sparse FEM systems require $\mathcal{O}(n^{3/2})$ operations in 2D and $\mathcal{O}(n^{2})$ in 3D~\cite{george1973nested,davis2016survey}, whereas the graph Laplacian system can in principle be solved in nearly linear 
time~\cite{spielman2004nearly}. This efficiency makes the framework well-suited for computationally demanding tasks such as large-scale network optimization.

We then applied this framework to study how redistributing material mass within a network affects energy dissipation. We first validated the method against analytical results for uniform 1D networks, confirming exact agreement. Our main focus in this work is analyzing the dissipative properties of networked materials in the plane of complex wavenumbers, $\xi(\omega)$. This helps in isolating the viscous (reflected in $\xi_r$) and elastic (reflected in $\xi_i$) properties of the materials and helps in uncovering lengthscale dependent design principles. After analysing dissipation profiles in bulk materials and establishing that $\xi_i$ and $\xi_r$ give wavelength and attenuation lengths of dissipation profiles respectively, we analyzed the relationship between dissipation and network structure in random networks.

We then showed, unsurprisingly, that random redistribution of cross-sectional areas is unlikely to improve upon uniform networks on average. However, though statistically rare, configurations with tapered geometries (thick rods near the source followed by progressively thinner ones) can exceed the uniform baseline dissipation in cases where driving frequency is close to certain normal modes of the network $(\xi_i=n\pi/L_{net}; n=1, 2,..)$.

Using gradient-based optimization under a fixed material cost, we showed that maximizing total energy dissipation yields nontrivial architecture. The optimal architectures resemble in some ways those of flow networks optimized for transport: both develop gradient structures with thick elements near the source that taper away from it. The correspondence is not one-to-one, however, because of differences in boundary conditions and cost functions. A detailed comparison of optimal solutions across the two settings is left to future work.

We also find that the optimal architecture becomes insensitive to boundary conditions imposed at the domain walls when the attenuation length is small compared to the system size. In this regime, mechanical energy is dissipated before stress waves reach the far boundaries, so the optimal mass distribution is determined entirely by the local material properties and source geometry rather than by the global boundary configuration.

Our results have direct implications for the rational design of architected materials for vibration isolation and protective equipment, particularly in the context of additive manufacturing where geometry and material distribution can be precisely controlled. Three concrete design principles emerge from this work: (a) energy dissipation in linear viscoelastic networks can be controlled by redistributing cross-sectional areas, without changing material composition; (b) the optimal mass distribution decays from the driven joint with a length scale set by the attenuation length of the base material; and (c)  for small attenuation lengths, the optimal architecture is independent of the imposed boundary conditions far away from source.

Several limitations of the present work should be noted. All results are obtained within a linear response framework appropriate for small amplitude dynamics, and computation of dissipation in networks 
approaching large displacements may not be quantitatively accurate due to the onset of nonlinear effects. Additionally, the current framework considers only axial deformation; incorporating bending elasticity would broaden the class of accessible architectures to include bending-dominated structures. Although the computational code is currently configured for 2D networks, extension to three-dimensional geometries is straightforward within the present formalism.

Looking forward, the design principles in large dissipative networks that have disorder in lattice sites and for general cases of $\xi_i$ need to be analysed rigorously and the current framework is ideally suited to address this problem. Our preliminary studies have shown that the optimization landscape becomes highly rugged with many resonant solutions requiring the use of stochastic optimization techniques. Apart from optimizing using gradient descent a particularly interesting open question is whether dissipative networks can adapt their architecture through local adaptation rules that can be autonomously implemented without local knowledge of global information. Bone remodeling offers a concrete template where trabecula thicken in proportion to experienced oscillatory strain~\cite{kumar2011dissipation, cowin2001bone, rubin1984regulation, hsieh2001effects, warden2004mechanotransduction, lanyon1982mechanically, hsieh2001effects}. Whether such locally adaptation converges to the gradient-optimized architectures identified here would connect our design principles to the broader question of mechanical adaptation in biological and engineered systems.

\begin{acknowledgments}
This research was funded by the Army Research Office (ARO) through the Multidisciplinary University Initiative (MURI) Grant No. W911NF2210219, the University of Pennsylvania Materials Research Science and Engineering Center (MRSEC) through Grant No. DMR-1720530 and DMR-2309043. PKP acknowledges support for this work through a seed grant from Penn's Materials Science and Engineering Center (MRSEC) grant DMR-1720530. NS acknowledges the use of large language models (Anthropic Claude) for coding and light editing.

\end{acknowledgments}

\bibliographystyle{apsrev4-2}
\bibliography{ref}

\end{document}


\title{Supplementary Information for:\\
Design principles for energy dissipation in viscoelastic network metamaterials}

\author{Niranjan Sarpangala}
\author{Sean Fancher}
\author{Prashant K. Purohit}
\author{Eleni Katifori}

\maketitle

\section*{Supplementary Information}

\section{Derivation of Network Laplacian}
We have the following constitutive relations for stress wave propagation in a single rod

\begin{subequations}
\begin{equation}
    \frac{\partial v}{\partial z}-\frac{\partial\epsilon}{\partial t} = 0,
    \label{matcon}
\end{equation}
\begin{equation}
    \frac{\partial\sigma}{\partial z}-\rho\frac{\partial v}{\partial t} = 0,
    \label{momcon}
\end{equation}
\label{coneqs}
\end{subequations}

where $v(z,t)$, $\epsilon(z,t)$, $\sigma(z,t)$ are the uniaxial velocity, strain and stress field respectively. The material density is denoted as $\rho$ and $z$ is a coordinate measured from one end of the rod. Here, we consider that the rods are viscoelastic, specifically, we consider the Boltzmann superposition principle for linear viscosity \cite{bland1960theory, hilton2017elastic, boltzmann1878theorie} given by the relation

\begin{equation}
    \epsilon\left(t\right) = \int_{0}^{t}d\tau\>c\left(t-\tau\right)\frac{d\sigma\left(\tau\right)}{d\tau},
    \label{Boltzdef}
\end{equation}

\noindent where $\epsilon$ is the strain of the material, $\sigma$ is the stress, and $c$ is the creep function. The time $t=0$ is chosen such that $\sigma(t)=0$ for all $t<0$. This formulation can be reexpressed slightly differently by additionally enforcing that $c(t)=0$ for all $t<0$ as well. These null regions of $\sigma(t)$ and $c(t)$ allow the bounds of integration to be freely expanded into $[-\infty , \infty]$.
This then allows us to denote the Fourier transform of $\epsilon$ as 

\begin{align}
    \tilde{\epsilon}\left(\omega\right) &= \frac{1}{2\pi}\int_{-\infty}^{\infty}dt\>\epsilon\left(t\right)e^{-i\omega t} = \frac{1}{2\pi}\int_{-\infty}^{\infty}dt\>e^{-i\omega t}\int_{-\infty}^{\infty}d\tau\>c\left(t-\tau\right)\frac{d\sigma\left(\tau\right)}{d\tau} \nonumber\\
    &= \frac{1}{2\pi}\int_{-\infty}^{\infty}d\tau\>e^{-i\omega\tau}\frac{d\sigma\left(\tau\right)}{d\tau}\int_{-\infty}^{\infty}dt\>e^{-i\omega\left(t-\tau\right)}c\left(t-\tau\right) = \tilde{c}\left(\omega\right)\int_{-\infty}^{\infty}d\tau\>e^{-i\omega\tau}\frac{d}{d\tau}\left(\int_{-\infty}^{\infty}d\omega'\>\tilde{\sigma}\left(\omega'\right)e^{i\omega'\tau}\right) \nonumber\\
    &= \tilde{c}\left(\omega\right)\int_{-\infty}^{\infty}d\omega'\>i\omega'\tilde{\sigma}\left(\omega'\right)\int_{-\infty}^{\infty}d\tau\>e^{i\tau\left(\omega'-\omega\right)} = 2\pi\tilde{c}\left(\omega\right)\int_{-\infty}^{\infty}d\omega'\>i\omega'\tilde{\sigma}\left(\omega'\right)\delta\left(\omega'-\omega\right) \nonumber\\
    &= 2\pi i\omega\tilde{c}\left(\omega\right)\tilde{\sigma}\left(\omega\right).
    \label{epFT}
\end{align}


With this, in Fourier space Eq \ref{coneqs} can be reexpressed as

\begin{subequations}
\begin{equation}
    \frac{\partial\tilde{v}}{\partial z}-i\omega\tilde{\epsilon} = \frac{\partial\tilde{v}}{\partial z}-2\pi\left(i\omega\right)^{2}\tilde{c}\left(\omega\right)\tilde{\sigma} = 0,
    \label{matconFT}
\end{equation}
\begin{equation}
    \frac{\partial\tilde{\sigma}}{\partial z}-i\omega\rho\tilde{v} = 0.
    \label{momconFT}
\end{equation}
\label{coneqsFT}
\end{subequations}

\noindent These can be readily solved to produce

\begin{subequations}
\begin{equation}
    \tilde{v}\left(z,\omega\right) = \frac{\xi\left(\omega\right)}{i\omega\rho}\left(\tilde{\sigma}_{+}\left(\omega\right)e^{z\xi\left(\omega\right)}-\tilde{\sigma}_{-}\left(\omega\right)e^{-z\xi\left(\omega\right)}\right),
    \label{sigsol_vel}
\end{equation}
\begin{equation}
    \tilde{\sigma}\left(z,\omega\right) = \tilde{\sigma}_{+}\left(\omega\right)e^{z\xi\left(\omega\right)}+\tilde{\sigma}_{-}\left(\omega\right)e^{-z\xi\left(\omega\right)},
    \label{sigsol_sig}
\end{equation}
\label{sigsols}
\end{subequations}

\noindent where $\tilde{\sigma}_{+}(\omega)$ and $\tilde{\sigma}_{-}(\omega)$ are determined by the boundary conditions and $\xi(\omega)$ is the dispersion function defined as

\begin{equation}
    \xi\left(\omega\right) = \sqrt{2\pi\left(i\omega\right)^{3}\rho\tilde{c}\left(\omega\right)}.
    \label{xidef}
\end{equation}

\noindent Eq. \ref{sigsols} can alternatively be written as

\begin{subequations}
\begin{equation}
    \tilde{v}\left(z,\omega\right) = \tilde{v}_{+}\left(\omega\right)e^{z\xi\left(\omega\right)}+\tilde{v}_{-}\left(\omega\right)e^{-z\xi\left(\omega\right)},
    \label{velsol_vel}
\end{equation}
\begin{equation}
    \tilde{\sigma}\left(z,\omega\right) = \frac{i\omega\rho}{\xi\left(\omega\right)}\left(\tilde{v}_{+}\left(\omega\right)e^{z\xi\left(\omega\right)}-\tilde{v}_{-}\left(\omega\right)e^{-z\xi\left(\omega\right)}\right).
    \label{velsol_sig}
\end{equation}
\label{velsols}
\end{subequations}

For the purposes of evaluating the dynamics of a rod of length $L$, the most useful forms of $\tilde{v}$ and $\tilde{\sigma}$ are given by

\begin{subequations}
\begin{equation}
    \tilde{v}\left(z,\omega\right) = \frac{\tilde{v}\left(0,\omega\right)\sinh\left(\left(L-z\right)\xi\left(\omega\right)\right)+ \allowbreak \tilde{v}\left(L,\omega\right)\sinh\left(z\xi\left(\omega\right)\right)}{\sinh\left(L\xi\left(\omega\right)\right)},
    \label{velbound_vel}
\end{equation}
\begin{equation}
    \tilde{\sigma}\left(z,\omega\right) = -\frac{i\omega\rho}{\xi\left(\omega\right)}\frac{\tilde{v}\left(0,\omega\right)\cosh\left(\left(L-z\right)\xi\left(\omega\right)\right)- \tilde{v}\left(L,\omega\right)\cosh\left(z\xi\left(\omega\right)\right)}{\sinh\left(L\xi\left(\omega\right)\right)}.
    \label{velbound_sig}
\end{equation}
\label{velbound}
\end{subequations}

\noindent From here we can consider an entire truss structure with connectivity laws given by

\begin{subequations}
\begin{equation}
    \hat{e}_{\mu\nu}^{\text{T}}\vec{w}_{\mu}\left(t\right) = v_{\mu\nu}\left(0,t\right),
    \label{wvelrel}
\end{equation}
\begin{equation}
    \vec{P}_{\mu}\left(t\right)+\sum_{\nu\in\mathcal{N}_{\mu}}\hat{e}_{\mu\nu}A_{\mu\nu}\sigma_{\mu\nu}\left(0,t\right) = \vec{0},
    \label{Psigrel}
\end{equation}
\label{trusscons}
\end{subequations}

\noindent where $\vec{w}_{\mu}$ is the velocity of joint $\mu$, $\vec{P}_{\mu}$ is the externally applied force at joint $\mu$, $\hat{e}_{\mu\nu}$ is the unit vector pointing from joint $\mu$ to joint $\nu$, and $A_{\mu\nu}$ is the cross sectional area of the rod connecting joints $\mu$ and $\nu$. By combining Eq. \ref{velbound_sig} with Eq. \ref{trusscons} we can readily produce the Fourier transformed forces at nodes, $\tilde{\vec{P}}_{\mu}(\omega)$

\begin{widetext}
\begin{align}
    \tilde{\vec{P}}_{\mu}\left(\omega\right) &= -\sum_{\nu\in\mathcal{N}_{\mu}}\hat{e}_{\mu\nu}A_{\mu\nu}\tilde{\sigma}_{\mu\nu}\left(0,\omega\right) \nonumber\\
    &= \sum_{\nu\in\mathcal{N}_{\mu}}\hat{e}_{\mu\nu}A_{\mu\nu}\frac{i\omega\rho_{\mu\nu}}{\xi_{\mu\nu}\left(\omega\right)}\frac{\tilde{v}_{\mu\nu}\left(0,\omega\right)\cosh\left(L_{\mu\nu}\xi_{\mu\nu}\left(\omega\right)\right)-\tilde{v}_{\mu\nu}\left(L_{\mu\nu},\omega\right)}{\sinh\left(L_{\mu\nu}\xi_{\mu\nu}\left(\omega\right)\right)} \nonumber\\
    &= \sum_{\nu\in\mathcal{N}_{\mu}}\hat{e}_{\mu\nu}A_{\mu\nu}\frac{i\omega\rho_{\mu\nu}}{\xi_{\mu\nu}\left(\omega\right)}\frac{\tilde{v}_{\mu\nu}\left(0,\omega\right)\cosh\left(L_{\mu\nu}\xi_{\mu\nu}\left(\omega\right)\right)+\tilde{v}_{\nu\mu}\left(0,\omega\right)}{\sinh\left(L_{\mu\nu}\xi_{\mu\nu}\left(\omega\right)\right)} \nonumber\\
    &= \sum_{\nu\in\mathcal{N}_{\mu}}\hat{e}_{\mu\nu}A_{\mu\nu}\frac{i\omega\rho_{\mu\nu}}{\xi_{\mu\nu}\left(\omega\right)}\frac{\hat{e}_{\mu\nu}^{\text{T}}\tilde{\vec{w}}_{\mu}\left(\omega\right)\cosh\left(L_{\mu\nu}\xi_{\mu\nu}\left(\omega\right)\right)+\hat{e}_{\nu\mu}^{\text{T}}\tilde{\vec{w}}_{\nu}\left(\omega\right)}{\sinh\left(L_{\mu\nu}\xi_{\mu\nu}\left(\omega\right)\right)}.
    \label{Pwrel_mu}
\end{align}
\end{widetext}

\noindent We define multi-vectors $\mathbf{\vec{P}}$ and $\mathbf{\vec{W}}$ of
N components each, where N is the number of joints in the network, with the $\mu^{th}$ components being the vectors $\vec{\tilde{P}}_{\mu}$ and $\vec{\tilde{w}}_{\mu}$  respectively. Based on above analysis, joint force and velocity multivectors $\mathbf{\vec{P}}$ and $\mathbf{\vec{W}}$ are related via

\begin{equation}
    \mathbf{\vec{P}} = \frac{1}{i\omega}\mathbf{\overset\Leftrightarrow{D}}\mathbf{\vec{W}} \quad\quad \implies \quad\quad \mathbf{\vec{P}} = \mathbf{\overset\Leftrightarrow{D}}\mathbf{\vec{U}},
    \label{multimatrel}
\end{equation}

\noindent where $\mathbf{\tilde{U}}$ is the joint displacement multivector of components N, where each component $\mu$ is the Fourier transformed displacement vector of joint $\mu$, $\vec{\tilde{u}}_{\mu}$. $\mathbf{\overset\Leftrightarrow{D}}$ is the network Laplacian, which has a $N\times N$ block matrix structure. The $\mu\nu^{th}$ element of $\mathbf{\overset\Leftrightarrow{D}}$ is a $d_{\mu}\times d_{\nu}$ matrix  ($d_{\mu}$ and $d_{\nu}$ are the dimensions needed to describe motions of joints $\mu$ and $\nu$ respectively) given by:

\begin{widetext}
\begin{equation}
    \mathbf{\overset\Leftrightarrow{D}}_{\mu\nu} = \begin{cases}
    \sum_{\gamma\in\mathcal{N}_{\mu}}\frac{\left(i\omega\right)^{2}\rho_{\mu\gamma}A_{\mu\gamma}}{\xi_{\mu\gamma}\left(\omega\right)}\coth\left(L_{\mu\gamma}\xi_{\mu\gamma}\left(\omega\right)\right)\hat{e}_{\mu\gamma}\hat{e}_{\mu\gamma}^{\text{T}} & \mu=\nu \\
    \frac{\left(i\omega\right)^{2}\rho_{\mu\nu}A_{\mu\nu}}{\xi_{\mu\nu}\left(\omega\right)}\text{csch}\left(L_{\mu\nu}\xi_{\mu\nu}\left(\omega\right)\right)\hat{e}_{\mu\nu}\hat{e}_{\nu\mu}^{\text{T}} & \nu\in\mathcal{N}_{\mu} \\
    \overset\Leftrightarrow{0} & \text{otherwise} \end{cases}
    \label{Dmndef}
\end{equation}
\end{widetext}

\section{Energy Dissipation}

The primary objective of this section will be simplify the rate of energy dissipation by a rod (Eq. 14 main text). Consider a rod with area $A$, density $\rho$ and length $L$. To get the rate of energy dissipated by the rod, we note that the mechanical power density, $\dot{\mathcal{W}}(z,t)$, will need to be multiplied by the rod cross sectional area (A), and averaged over one full period to get the dissipation rate per unit length which is then integrated over the length of the rod (L). Mathematically, this process takes the form

\begin{align}
    Q &= \frac{A}{T}\int_{0}^{L}dz\int_{0}^{T}dt\>\dot{\mathcal{W}}\left(z,t\right) = -\frac{A}{T}\int_{0}^{L}dz\int_{0}^{T}dt\left(\sum_{n=-\infty}^{\infty}\tilde{\sigma}_{n}\left(z\right)e^{2\pi int/T}\right)\frac{\partial}{\partial t}\left(\sum_{m=-\infty}^{\infty}\tilde{\epsilon}_{m}\left(z\right)e^{2\pi imt/T}\right) \nonumber\\
    &= -\frac{A}{T}\sum_{n,m=-\infty}^{\infty}\int_{0}^{L}dz\int_{0}^{T}dt\>\frac{2\pi im}{T}\tilde{\sigma}_{n}\left(z\right)\tilde{\epsilon}_{m}\left(z\right)e^{2\pi i\left(n+m\right)t/T} = -\frac{2\pi iA}{T}\sum_{n,m=-\infty}^{\infty}\int_{0}^{L}dz\>m\tilde{\sigma}_{n}\left(z\right)\tilde{\epsilon}_{m}\left(z\right)\delta_{n,-m}.
    \label{Qfulldef}
\end{align}

From here, the $m$ summation becomes trivially easy due to the Kronecker $\delta$ symbol and will simply cause $m$ to be replaced by $-n$. We can also note the conjugate-negation relation $\tilde{\epsilon}_{-n}(z)=\tilde{\epsilon}_{n}^{*}(z)$ and utilize Eq. \ref{epFT} (with the understanding that $\omega$ is replaced by $2\pi n/T$) to further reduce Eq. \ref{Qfulldef} to

\begin{align}
    Q &= -\frac{2\pi iA}{T}\sum_{n=-\infty}^{\infty}\int_{0}^{L}dz\>\left(-n\right)\tilde{\sigma}_{n}\left(z\right)\tilde{\epsilon}_{n}^{*}\left(z\right) = \frac{2\pi iA}{T}\sum_{n=-\infty}^{\infty}\int_{0}^{L}dz\>n\tilde{\sigma}_{n}\left(z\right)\left(\frac{\left(2\pi\right)^{2}in}{T}\tilde{c}_{n}\tilde{\sigma}_{n}\left(z\right)\right)^{*} \nonumber\\
    &= \frac{\left(2\pi\right)^{3}A}{T^{2}}\sum_{n=-\infty}^{\infty}\int_{0}^{L}dz\>n^{2}\left|\tilde{\sigma}_{n}\left(z\right)\right|^{2}\tilde{c}_{n}^{*} = \frac{2\left(2\pi\right)^{3}A}{T^{2}}\sum_{n=1}^{\infty}n^{2}\text{Re}\left(\tilde{c}_{n}\right)\int_{0}^{L}dz\>\left|\tilde{\sigma}_{n}\left(z\right)\right|^{2},
    \label{Qepred}
\end{align}

\noindent where in the last equality we have applied the conjugate-negation relation to $\tilde{\sigma}_{n}(z)$ and $\tilde{c}_{n}$ and assumed the $n=0$ term vanishes due to the $n^{2}$ factor overpowering any divergences that may occur in $\tilde{\sigma}_{0}(z)$ or $\tilde{c}_{0}$.

We next isolate the $z$ integral from Eq. \ref{Qepred}. Assuming known velocity boundary conditions, we can use Eq. \ref{velbound_sig} to expand this as

\begin{align}
    &\int_{0}^{L}dz\>\left|\tilde{\sigma}_{n}\left(z\right)\right|^{2} = \int_{0}^{L}dz\>\left(\frac{2\pi\rho}{T\left|\xi_{n}\right|}\right)^{2}\left|\frac{\tilde{v}_{n}\left(0\right)\cosh\left(\left(L-z\right)\xi_{n}\right)-\tilde{v}_{n}\left(L\right)\cosh\left(z\xi_{n}\right)}{\sinh\left(L\xi_{n}\right)}\right|^{2}\nonumber\\
    &= \left(\frac{2\pi\rho}{T\left|\xi_{n}\sinh\left(L\xi_{n}\right)\right|}\right)^{2}\int_{0}^{L}dz\>\Bigl(\left|\tilde{v}_{n}\left(0\right)\right|^{2}\cosh\left(\left(L-z\right)\xi_{n}\right)\cosh\left(\left(L-z\right)\xi_{n}^{*}\right)+\left|\tilde{v}_{n}\left(L\right)\right|^{2}\cosh\left(z\xi_{n}\right)\cosh\left(z\xi_{n}^{*}\right) \nonumber\\
    &\quad\quad\quad\quad\quad\quad\quad\quad\quad\quad\quad\quad\quad -\tilde{v}_{n}\left(0\right)\tilde{v}_{n}^{*}\left(L\right)\cosh\left(\left(L-z\right)\xi_{n}\right)\cosh\left(z\xi_{n}^{*}\right)-\tilde{v}_{n}^{*}\left(0\right)\tilde{v}_{n}\left(L\right)\cosh\left(\left(L-z\right)\xi_{n}^{*}\right)\cosh\left(z\xi_{n}\right)\Bigr) \nonumber\\
    &= \left(\frac{2\pi\rho}{T\left|\xi_{n}\sinh\left(L\xi_{n}\right)\right|}\right)^{2}\left(\left(\left|\tilde{v}_{n}\left(0\right)\right|^{2}+\left|\tilde{v}_{n}\left(L\right)\right|^{2}\right)\frac{\xi_{n}\sinh\left(L\xi_{n}\right)\cosh\left(L\xi_{n}^{*}\right)-\xi_{n}^{*}\sinh\left(L\xi_{n}^{*}\right)\cosh\left(L\xi_{n}\right)}{\left(\xi_{n}\right)^{2}-\left(\xi_{n}^{*}\right)^{2}}\right. \nonumber\\
    &\quad\quad\quad\quad\quad\quad\quad\quad\quad\quad\quad\quad \left.-\left(\tilde{v}_{n}\left(0\right)\tilde{v}_{n}^{*}\left(L\right)+\tilde{v}_{n}^{*}\left(0\right)\tilde{v}_{n}\left(L\right)\right)\frac{\xi_{n}\sinh\left(L\xi_{n}\right)-\xi_{n}^{*}\sinh\left(L\xi_{n}^{*}\right)}{\left(\xi_{n}\right)^{2}-\left(\xi_{n}^{*}\right)^{2}}\right) \nonumber\\
    &= \left(\frac{2\pi\rho}{T}\right)^{2}\frac{L^{3}}{\text{Im}\left(\left(L\xi_{n}\right)^{2}\right)}\left(\left(\left|\tilde{v}_{n}\left(0\right)\right|^{2}+\left|\tilde{v}_{n}\left(L\right)\right|^{2}\right)\text{Im}\left(\frac{\coth\left(L\xi_{n}^{*}\right)}{L\xi_{n}^{*}}\right)-2\text{Re}\left(\tilde{v}_{n}\left(0\right)\tilde{v}_{n}^{*}\left(L\right)\right)\text{Im}\left(\frac{\text{csch}\left(L\xi_{n}^{*}\right)}{L\xi_{n}^{*}}\right)\right).
    \label{sigmasqint}
\end{align}

\noindent From here we can additionally use Eq. \ref{xidef} to note that

\begin{equation}
    \text{Im}\left(\left(L\xi_{n}\right)^{2}\right) = \text{Im}\left(2\pi L^{2}\left(\frac{2\pi in}{T}\right)^{3}\rho\tilde{c}_{n}\right) = -\frac{\left(2\pi\right)^{4}n^{3}L^{2}\rho}{T^{3}}\text{Re}\left(\tilde{c}_{n}\right).
    \label{Imxirel}
\end{equation}

\noindent We can also incorporate the resulting negative sign into the other imaginary operators found in Eq. \ref{sigmasqint} via the relation $-\text{Im}(z^{*})=\text{Im}(z)$ to produce

\begin{equation}
    \int_{0}^{L}dz\>\left|\tilde{\sigma}_{n}\left(z\right)\right|^{2} = \frac{LT\rho}{\left(2\pi\right)^{2}n^{3}\text{Re}\left(\tilde{c}_{n}\right)}\left(\left(\left|\tilde{v}_{n}\left(0\right)\right|^{2}+\left|\tilde{v}_{n}\left(L\right)\right|^{2}\right)\text{Im}\left(\frac{\coth\left(L\xi_{n}\right)}{L\xi_{n}}\right)-2\text{Re}\left(\tilde{v}_{n}\left(0\right)\tilde{v}_{n}^{*}\left(L\right)\right)\text{Im}\left(\frac{\text{csch}\left(L\xi_{n}\right)}{L\xi_{n}}\right)\right).
    \label{sigmasqint_red}
\end{equation}

\noindent Finally, substituting Eq. \ref{sigmasqint_red} into Eq. \ref{Qepred} yields

\begin{equation}
    Q = \frac{4\pi AL\rho}{T}\sum_{n=1}^{\infty}\frac{1}{n}\left(\left(\left|\tilde{v}_{n}\left(0\right)\right|^{2}+\left|\tilde{v}_{n}\left(L\right)\right|^{2}\right)\text{Im}\left(\frac{\coth\left(L\xi_{n}\right)}{L\xi_{n}}\right)-2\text{Re}\left(\tilde{v}_{n}\left(0\right)\tilde{v}_{n}^{*}\left(L\right)\right)\text{Im}\left(\frac{\text{csch}\left(L\xi_{n}\right)}{L\xi_{n}}\right)\right).
    \label{Qfinal}
\end{equation}

\section{Note on global standing waves}
\label{sec:standing_wave}

The imaginary part of the dispersion function, $\xi_i$, sets a
wavelength $\lambda = \pi/|\xi_i|$ for spatial oscillations of stress
and velocity along a rod. Particular values of $\xi_i$ correspond to
standing-wave resonances of the network as a whole, and we derive that
condition here.

Consider a network of total length $L_{\text{net}}$ along the direction
of excitation, driven by a prescribed harmonic displacement at $z=0$
and held fixed at $z=L_{\text{net}}$. If we have a uniform network and excitation is along the horizontal axis, the entire network can be treated as a
single effective rod of length $L_{\text{net}}$ for the purposes of
identifying its global standing-wave modes. The velocity field then
takes the form
\begin{equation}
  \tilde{v}(z,\omega) \;=\; \tilde{v}(0,\omega)
    \frac{\sinh\bigl((L_{\text{net}}-z)\xi(\omega)\bigr)}
         {\sinh\bigl(L_{\text{net}}\xi(\omega)\bigr)}
  + \tilde{v}(L_{\text{net}},\omega)
    \frac{\sinh\bigl(z\,\xi(\omega)\bigr)}
         {\sinh\bigl(L_{\text{net}}\xi(\omega)\bigr)}.
  \label{eq:si_vfield}
\end{equation}
With $\tilde{v}(L_{\text{net}},\omega) = 0$ and $\tilde{v}(0,\omega)$
finite and nonzero, the response amplitude diverges whenever
\begin{equation}
  \sinh\bigl(L_{\text{net}}\,\xi(\omega)\bigr) \;=\; 0.
  \label{eq:si_resonance_condition}
\end{equation}
These divergences are the resonant modes of the system: at these
$\xi$, a finite drive produces an infinite response, signaling a
lossless eigenmode of the underlying homogeneous boundary-value
problem.

In the lossless limit $\xi_r \to 0$, the dispersion function becomes
purely imaginary, $\xi \to i\xi_i$, and
\begin{equation}
  \sinh(L_{\text{net}}\,\xi) = i\,\sin(L_{\text{net}}\xi_i).
\end{equation}
The resonance condition Eq.~(\ref{eq:si_resonance_condition}) then
reduces to $\sin(L_{\text{net}}\xi_i) = 0$, giving the standing-wave
spectrum
\begin{equation}
  \xi_i \;=\; \frac{n\pi}{L_{\text{net}}},
  \qquad n = 1, 2, 3, \ldots
  \label{eq:si_standing_wave}
\end{equation}

For weakly dissipative materials, the response
amplitude does not diverge but remains sharply peaked near
$\xi_i = n\pi/L_{\text{net}}$. The simulations in main text operate in this near-resonant regime, with
$L_{\text{net}} = 10$ and $\xi_i = 3.15$. (We are slightly off the
$n=10$ resonance at $\xi_i = \pi$, chosen this way to avoid the
numerical singularity of the exact eigenmode while remaining
well within its peak. At this specific value $\xi_i = \pi $ each individual rod of
length $L = 1$ is also near its own single rod resonance.

\subsection{Asymptotic behavior of the dissipation kernel near resonance}

For the above boundary condition, the dissipated power $Q$ derived in Eq.~(\ref{Qfinal}) depends only on the kernel $\mathrm{Im}[\coth(L_{net}\xi)/(L_{net}\xi)]$ when $v(L_{net})=0$. We here evaluate its asymptotic behavior as $\xi_r \to 0$ at fixed $\xi_i = n\pi/L_{net}$, which corresponds to approaching the $n$-th standing-wave resonance from the dissipative side. Without loss of generality we can set $L_{net}=1$ and consider $\xi = \xi_r + i n\pi$.

Using $\sinh(\xi_r + in\pi) = (-1)^n \sinh(\xi_r)$ and $\cosh(\xi_r + in\pi) = (-1)^n \cosh(\xi_r)$, we have
\begin{equation}
    \coth(\xi_r + in\pi) = \coth(\xi_r).
\end{equation}
In the small-$\xi_r$ limit, $\coth(\xi_r) = 1/\xi_r + \xi_r/3 + \mathcal{O}(\xi_r^3)$, so
\begin{equation}
    \frac{\coth(\xi_r + in\pi)}{\xi_r + in\pi}
    \;\approx\; \frac{1}{\xi_r (\xi_r + in\pi)}
    \;=\; \frac{\xi_r - in\pi}{\xi_r(\xi_r^2 + n^2\pi^2)}.
\end{equation}
Taking the imaginary part,
\begin{equation}
    \mathrm{Im}\!\left[\frac{\coth(\xi_r + in\pi)}{\xi_r + in\pi}\right]
    \;=\; -\frac{n\pi}{\xi_r(\xi_r^2 + n^2\pi^2)}.
\end{equation}
For $\xi_r \ll n\pi$, the $\xi_r^2$ term in the denominator is negligible, yielding
\begin{equation}
    \mathrm{Im}\!\left[\frac{\coth(\xi_r + in\pi)}{\xi_r + in\pi}\right]
    \;\xrightarrow[\xi_r \to 0]{}\; -\frac{1}{n\pi\,\xi_r}.
    \label{eq:kernel_asymptote}
\end{equation}
Thus the kernel diverges as $-1/(\xi_r)$ as $\xi_r\rightarrow0$. 



\section*{References}

\bibliographystyle{apsrev4-2}
\bibliography{ref}
\newpage

\section*{Supplementary Figures}

\begin{figure}[ht]
  \centering
  \includegraphics[width=0.6\textwidth]{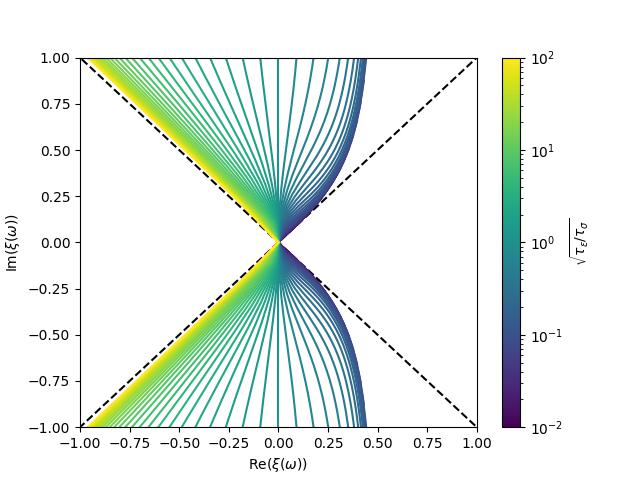}
  \caption{Dispersion function $\xi(\omega)$ for the standard
  linear solid model. Real and imaginary parts of $\xi(\omega)$
  computed from Eq.~8 of the main text as a function
  of driving frequency $\omega$, for range of values of $\sqrt{\tau_{\epsilon}/\tau_{\sigma}}$. Across the full frequency range,
  $|\xi_i(\omega)| > |\xi_r(\omega)|$, constraining physically
  realizable operating points to the wedge $|\xi_r| < |\xi_i|$ in the
  complex-$\xi$ plane.}
  \label{fig:si_xi_omega}
\end{figure}

\begin{figure*}
    \centering
    \includegraphics[width=0.5\linewidth]{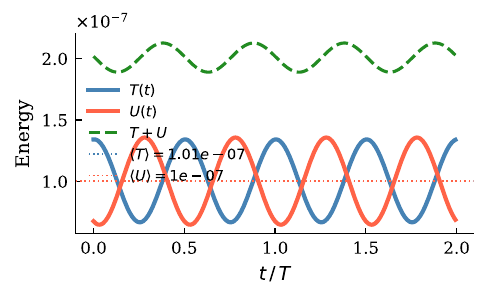}
    
    \caption{Energy balance for a single viscoelastic rod. The rod is clamped at $z=L$ and driven harmonically at $z=0$. Parameters are the same as Fig.~2: $L=10$, $A=0.1$, $\rho=1$, $\omega=1$, $u_0=10^{-3}$. Complex wavenumber, $\xi=-0.14+3.15i$.  Kinetic energy $T(t) = \int_0^L \tfrac{1}{2}\rho A\,\dot{u}^2\,dz$ and elastic energy $U(t) = \int_0^L \tfrac{1}{2}E'A\,\varepsilon^2\,dz$ computed from the analytical frequency-domain solution (Eq.~3, main text), with the spatial integral evaluated numerically. ($E'$ is the storage modulus which is the real part of complex modulus, $E' = \mathrm{Re}[-\xi^2/\rho\omega^2]$). Both oscillate at twice the driving frequency $2\omega$; their sum $T+U$ is strictly periodic with no net drift, confirming that no energy accumulates in the rod over a cycle. The cycle-averaged power delivered by the boundary force, $\langle P_\mathrm{in}\rangle = \langle{-A\,\sigma(0,t)\,\dot{u}(0,t)}\rangle$, was computed from the analytical solution (Eq.~3, main text) and equals $1.787\times10^{-8}$. The total energy dissipated per unit time, $Q_\mathrm{01}$ computed as in Fig.~2 and equals $1.787\times10^{-8}$. Because $T+U$ is strictly periodic, $\langle d(T+U)/dt\rangle = 0$, and hence $\langle P_\mathrm{in}\rangle = Q_\mathrm{01}$, as expected.}
    \label{fig:energy_balance}
\end{figure*}

\begin{figure*}
    \centering
    \includegraphics[width=\linewidth]{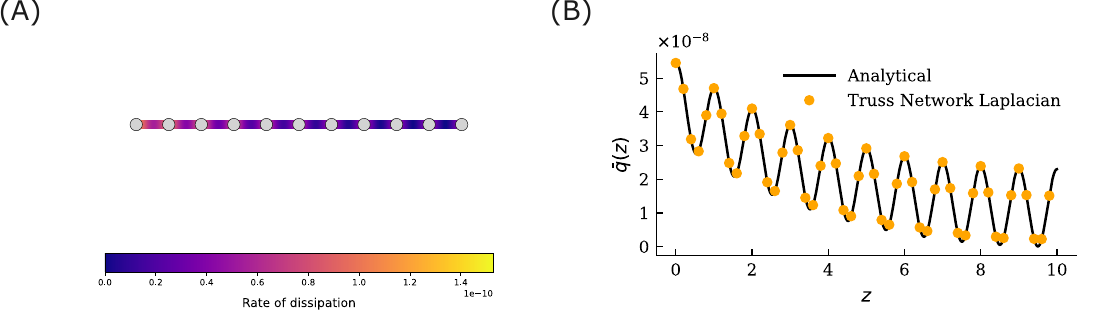}
    \caption{Energy dissipation profile in uniform network for benchmarking. Rate of energy dissipation per unit length $\bar{q}(z)$ in a uniform network with 10 rods, rightmost joint is fixed, harmonic displacement is provided at the leftmost joint. $\xi=-0.14 +3.15i$}
    \label{fig:total_energy_1d_network}
\end{figure*}

\begin{figure*}
    \centering
    \includegraphics[width=0.75\linewidth]{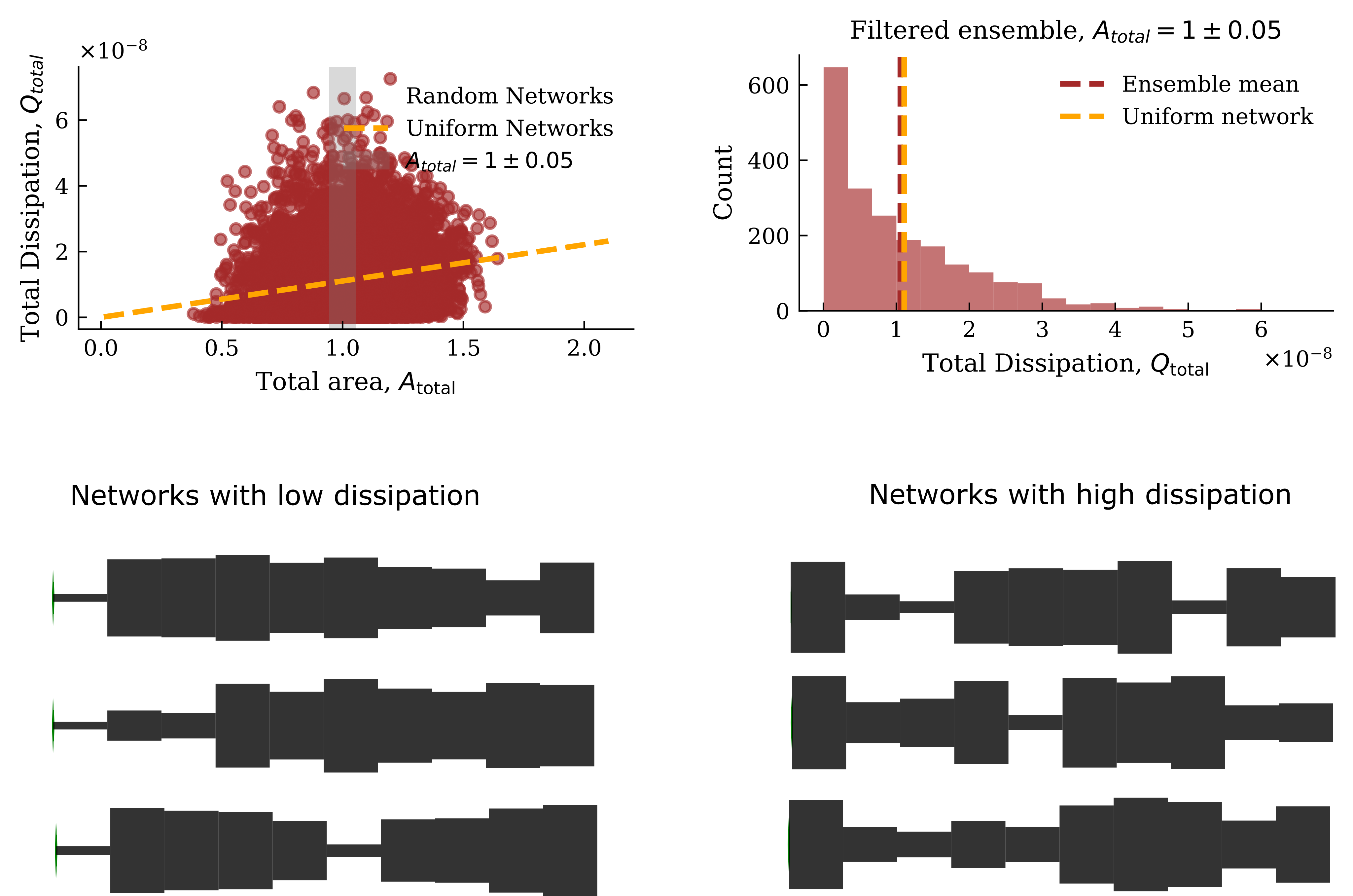}
    \caption{Analysis of 1D random networks for $\xi=-0.14 +5i$ (Rest of the parameters and boundary conditions are same as Fig 4 of main text)}
    \label{fig:1d_random_network_general_case}
\end{figure*}
\begin{figure*}
    \centering
    \includegraphics[width=0.75\linewidth]{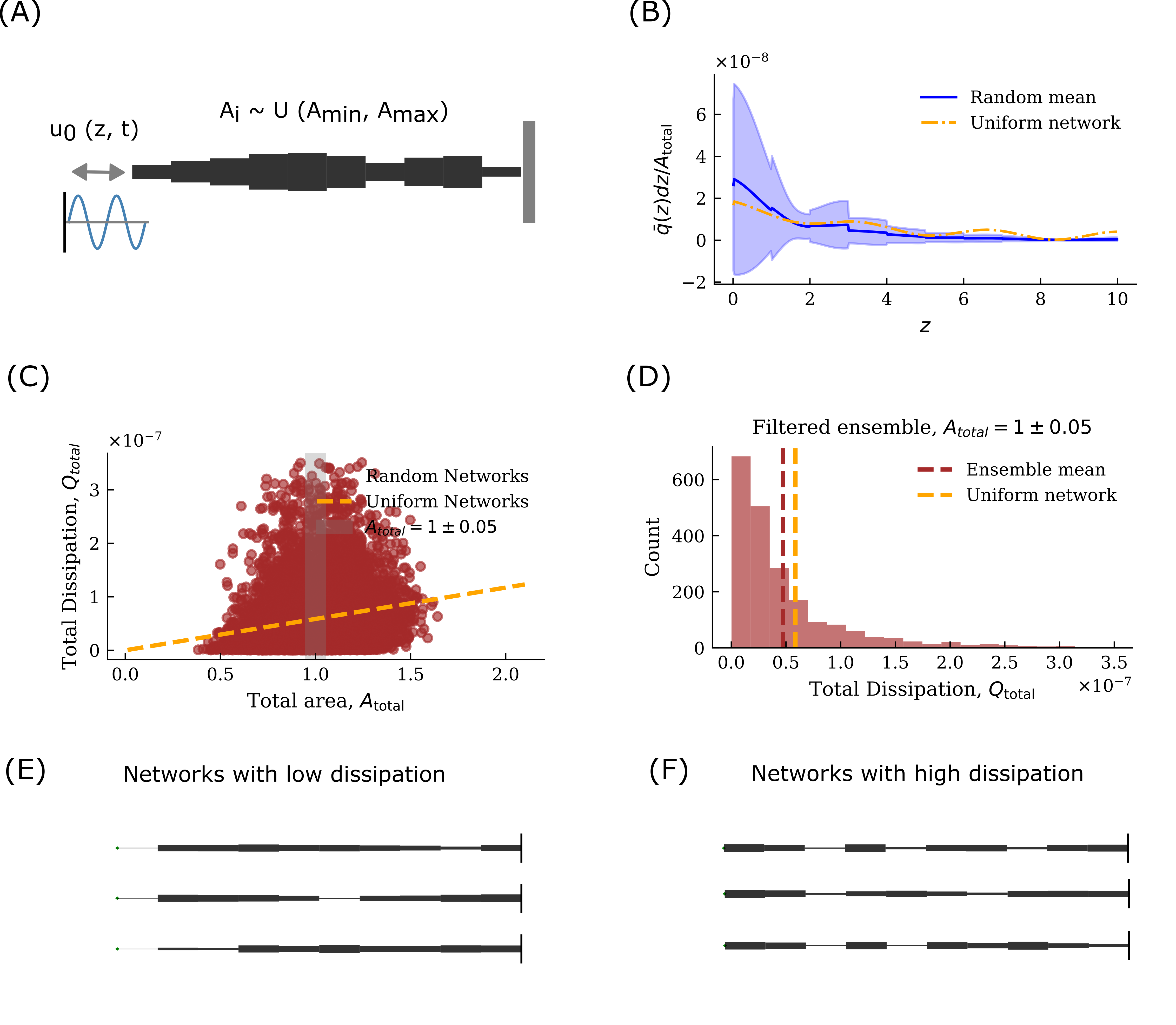}
    \caption{Analysis of 1D random networks for $\xi=-0.14 +3\pi /10 \, i$ (Rest of the parameters and boundary conditions are same as Fig 4 of main text)}
    \label{fig:1d_random_network_3pi_lnet}
\end{figure*}

\begin{figure*}
    \centering
    \includegraphics[width=\linewidth]{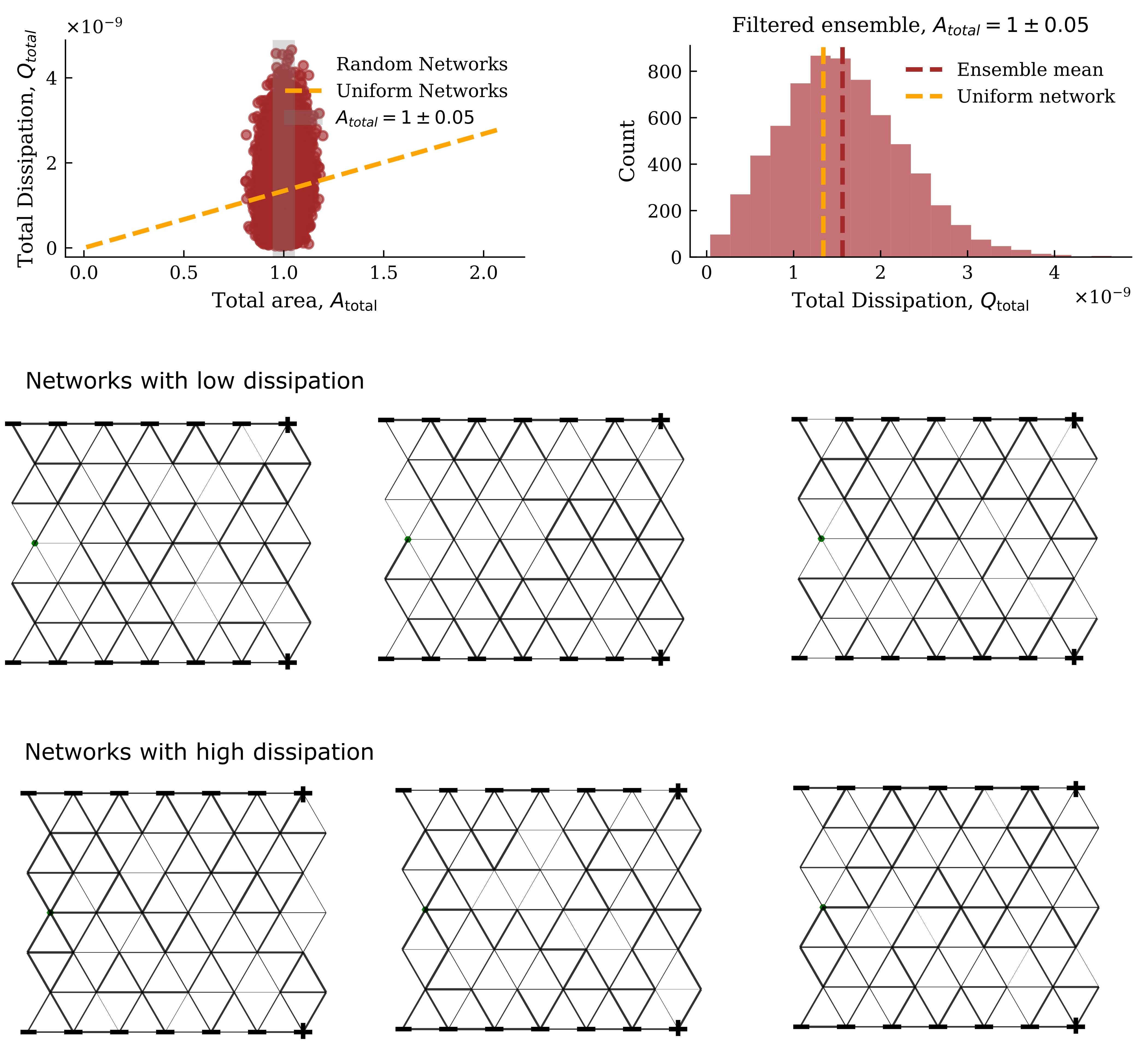}
    \caption{Analysis of 2D random networks for $\xi=-0.14 +5i$ (Rest of the parameters and boundary conditions are same as Fig 5 of main text)}
    \label{fig:1d_random_network_general_case}
\end{figure*}

\begin{figure*}
    \centering
    \includegraphics[width=0.5\linewidth]{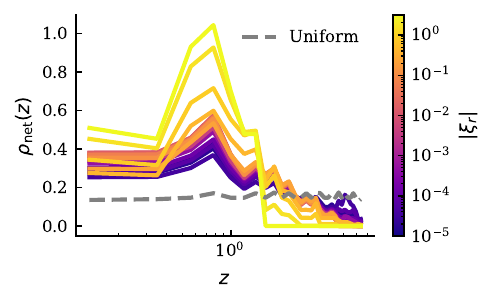}
    \caption{Projected linear mass density $\rho_{\rm net}(z)$ of optimized networks as a function of horizontal position $z$, for a 7×7 triangular lattice ($L=1$, $C_{target}=1$). Each curve is the trial-averaged profile over $N_{\rm trials}=10$ independent optimizations; color encodes $|\xi_r|$ on a logarithmic scale (20 values spanning $|\xi_r| \in [10^{-5}, 3.14]$). The dashed grey curve shows the uniform reference network. Data from optimization in Fig.~7 of main text.}
\end{figure*}

\begin{figure*}
    \centering
    \includegraphics[width=0.5\linewidth]{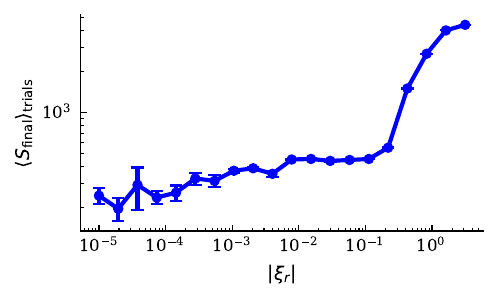}
    \caption{Mean nearest-neighbor area correlation $\langle S \rangle_\mathrm{trials}$ in optimized networks as a function of $\xi_r$. Points denote the mean across trials; error bars represent the standard deviation. $S = \left\langle \frac{(A_{ij} - \langle A \rangle)(A_{ik} - \langle A \rangle)}{\langle A \rangle^2} \right\rangle $. Data from optimization in Fig. 7 of main text.}
\end{figure*}

\begin{figure*}
    \centering
    \includegraphics[width=0.5\linewidth]{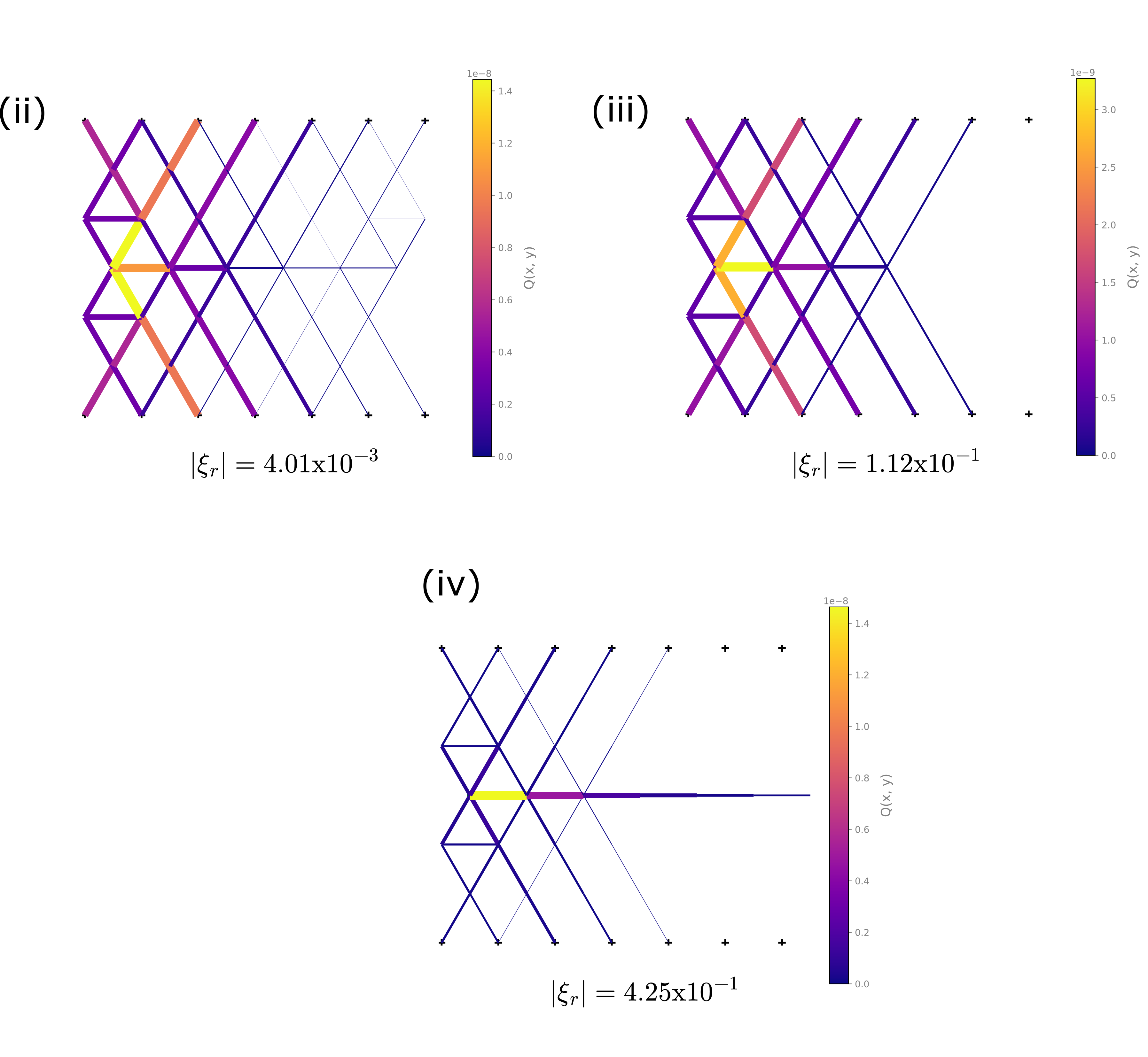}
    \caption{Representative optimized network architectures at selected values of $|\xi_r|$ for a boundary condition where joints in the top and bottom rows are fixed in position along x and y axes. Rod thickness is proportional to the square root of cross-sectional area relative to maximum area in the network, and color indicates energy dissipated by rods. Joints on the top and bottom row are fixed as indicated by `$+$' symbol. Network parameters: $\xi_i=3.15$, $\rho=1$, $L=1$ for all rods in networks. Optimization parameters: $\alpha= 500$, $\alpha_{min}=0.05$, $l_f=0.1$, $N_{plateau}=200$, $\epsilon_{rel}=10^{-6}$, $C_{target}=1$. Colorbar range is set independently per panel.}
\end{figure*}

\begin{figure*}
    \centering
    \includegraphics[width=0.65\linewidth]{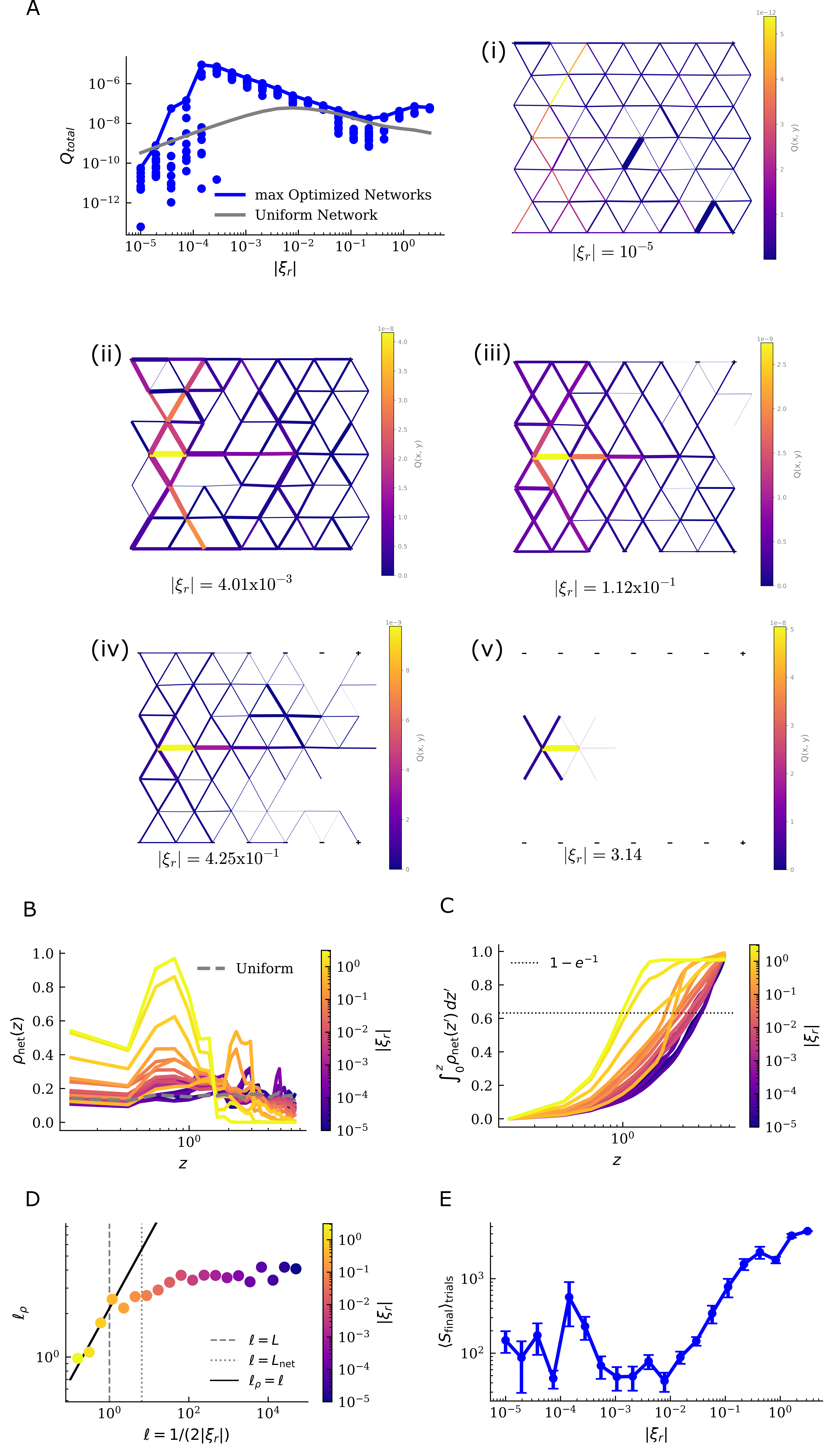}
    \caption{Robustness of the $\ell_\rho$–$\ell$ scaling relation under geometric perturbation ($\sigma = 0.02$). The triangular lattice node positions were displaced by independent Gaussian noise with standard deviation $\sigma = 0.02$ (in units of the lattice spacing) before optimization. Panels show the same analysis as Fig. 7\&8 of main text: \textbf{(A)} (top left) $Q_\text{total}$ across optimized networks versus $|\xi_r|$; (i-v) Representative optimized network geometries at selected $|\xi_r|$ values. \textbf{(B)} Density $\rho_\text{net}(z)$ and \textbf{(C)} Its cumulative integral; \textbf{(D)} Dissipation length $\ell_\rho$ versus $\ell = 1/(2|\xi_r|)$, and \textbf{(E)} Mean correlation $\langle S_\text{final}\rangle_\text{trials}$ versus $|\xi_r|$. The $\ell_\rho \propto \ell$ scaling and the transition in optimized network topology are preserved under perturbation. All parameters are same as Fig.~7 in main text.}
\end{figure*}

\begin{figure*}
    \centering
    \includegraphics[width=0.65\linewidth]{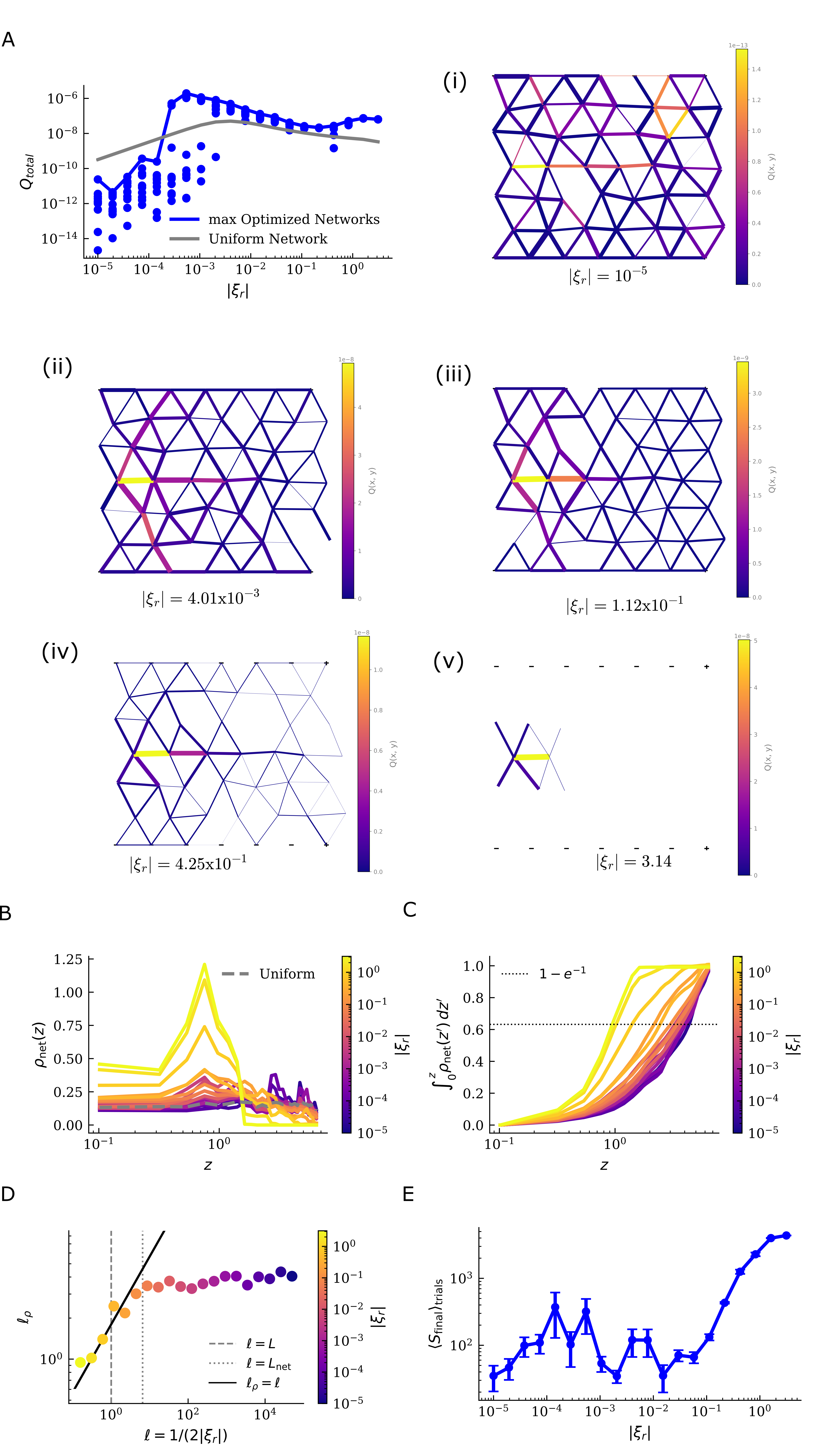}
    \caption{Robustness of the $\ell_\rho$–$\ell$ scaling relation under geometric perturbation ($\sigma = 0.10$). The triangular lattice node positions were displaced by independent Gaussian noise with standard deviation $\sigma = 0.10$ (in units of the lattice spacing) before optimization. Panels show the same analysis as Fig. 7\&8 of main text: \textbf{(A)} (top left) $Q_\text{total}$ across optimized networks versus $|\xi_r|$; (i-v) Representative optimized network geometries at selected $|\xi_r|$ values. \textbf{(B)} Density $\rho_\text{net}(z)$ and \textbf{(C)} Its cumulative integral; \textbf{(D)} Dissipation length $\ell_\rho$ versus $\ell = 1/(2|\xi_r|)$, and \textbf{(E)} Mean correlation $\langle S_\text{final}\rangle_\text{trials}$ versus $|\xi_r|$. The $\ell_\rho \propto \ell$ scaling and the transition in optimized network topology are preserved under perturbation. All parameters are same as Fig.~7 in main text.}
\end{figure*}